\documentclass[USenglish,oneside,twocolumn]{article}

\usepackage[utf8]{inputenc}
\usepackage[big]{dgruyter_NEW}
\usepackage{epsfig,endnotes}
\usepackage{varwidth}
\usepackage{algpseudocode}
\usepackage{balance}
\usepackage{color}
\usepackage{nicefrac}
\usepackage{amsmath}
\usepackage{bm}
\usepackage{mathtools}
\usepackage{multirow}
\usepackage{bigdelim}
\usepackage{mathtools}
\usepackage{amssymb}
\usepackage{indentfirst}
\usepackage{booktabs}
\usepackage{enumitem}
\usepackage{float}
\usepackage{graphicx}
\usepackage{caption, subcaption}
\usepackage[normalem]{ulem}
\usepackage[titletoc]{appendix}
\usepackage{flushend}

\newtheorem{theorem}{Theorem}
\newtheorem{lemma}{Lemma}
\newcommand{\minus}{\scalebox{0.75}[1.0]{$-$}}

\newcommand{\ourprotocol}{Root ORAM}
\newcommand{\etal}{\textit{et al}.}
\cclogo{\includegraphics{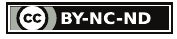}}

\makeatletter
\let\origsection\section
\renewcommand\section{\@ifstar{\starsection}{\nostarsection}}

\newcommand\nostarsection[1]
{\sectionprelude\origsection{#1}\sectionpostlude}

\newcommand\starsection[1]
{\sectionprelude\origsection*{#1}\sectionpostlude}

\newcommand\sectionprelude{%
  \vspace{-1em}
}

\newcommand\sectionpostlude{%
  \vspace{-1em}
}
\makeatother

\makeatletter
\let\origsubsection\subsection
\renewcommand\subsection{\@ifstar{\starsubsection}{\nostarsubsection}}
\newcommand\nostarsubsection[1]
{\subsectionprelude\origsubsection{#1}\subsectionpostlude}
\newcommand\starsubsection[1]
{\subsectionprelude\origsubsection*{#1}\subsectionpostlude}
\newcommand\subsectionprelude{%
  \vspace{-1.5em}}
\newcommand\subsectionpostlude{%
  \vspace{-1em}}
\makeatother

 \graphicspath{{./Images/}}
\DeclareGraphicsExtensions{.pdf,.jpg,.png}
\newtheorem{defn}{Definition}

\begin{document}
 
  \author*[1]{Sameer Wagh}
  \author[2]{Paul Cuff}
  \author[3]{Prateek Mittal}
  \affil[1]{Princeton University, E-mail: swagh@princeton.edu}
  \affil[2]{Renaissance Technonogies, E-mail: paul.cuff@gmail.com}
  \affil[3]{Princeton University, E-mail: pmittal@princeton.edu}

\title{\huge Differentially Private Oblivious RAM}
\runningtitle{Differentially Private Oblivious RAM}

\begin{abstract}
{In this work, we investigate if statistical privacy can 
enhance the performance of ORAM mechanisms while providing 
rigorous privacy guarantees. We propose a formal and rigorous framework for developing ORAM protocols with statistical security viz., a \textit{differentially private ORAM} (DP-ORAM).
We present \ourprotocol{}, a family of DP-ORAMs that provide a tunable, multi-dimensional 
trade-off between the desired bandwidth overhead, local storage and system security.\\
We theoretically analyze \ourprotocol{} to quantify both its security and performance.
We experimentally demonstrate the benefits of \ourprotocol{} and find that (1) \ourprotocol{} can reduce local storage overhead by about $2\times$ for a reasonable values of privacy budget, significantly enhancing performance in memory limited platforms such as trusted execution environments, and (2) \ourprotocol{} allows tunable trade-offs between bandwidth, storage, and privacy, reducing bandwidth overheads by up to $2\times$-$10\times$ (at the cost of increased storage/statistical privacy), enabling significant reductions in ORAM access latencies for cloud environments. We also analyze the privacy guarantees of DP-ORAMs through the lens of information theoretic metrics of Shannon entropy and Min-entropy~\cite{paulDP}. Finally, \ourprotocol{} is ideally suited for applications which have a similar access pattern, and we showcase its utility via the application of Private Information Retrieval.}
\end{abstract}


	\journalname{Proceedings on Privacy Enhancing Technologies}
	\DOI{10.1515/popets-2018-0032}
	\startpage{64}
 	\received{2018-02-28}
	\revised{2018-06-15}
	\accepted{2018-06-16}

	\journalyear{}
	\journalvolume{2018}
	\journalissue{4}

\maketitle

\vspace{-3.5em}
\section{Introduction}\label{sec:introduction}

Oblivious RAM (ORAM), first introduced by Goldreich 
and Ostrovsky~\cite{goldreichoram, goldreich1987towards}, 
is a cryptographic primitive which allows a client to 
protect its data access pattern from an untrusted server 
storing the data. Since its introduction, substantial 
progress has been made by the research community in 
developing novel and efficient 
ORAM schemes~\cite{pathoram,ringoram,gentryoramSC,burstoram,oblivistore, SokDatabaseSearch, tree_based_orams, ren2013design}. 
Recent work has also shown the promise of using ORAMs as a 
critical component in developing protocols for 
Secure Multi-Party Computation~\cite{gentryoramSC}.

ORAMs can mitigate side-channel attacks~\cite{accesspatterndisclosure,dautrich2013compromising} in two typical deployment contexts:
(1) Trusted Execution Environments such as SGX-based enclaves~\cite{sgxreference}, involving communications between last-level cache (LLC) and DRAM, and 
(2) Client-server environments, such as communications between smartphones and cloud servers. However, a key bottleneck in 
the practical deployment of ORAM protocols in these contexts is the performance overhead.  
For instance, even the most efficient ORAM protocols~\cite{ringoram,pathoram,burstoram,SSSoram} incur a 
logarithmic bandwidth overhead $($>$100 \times$-$200 \times)$ as well as a logarithmic local storage/stash overhead\footnote{Recent work such as Circuit ORAM~\cite{circuitoram} require constant local memory but increase the protocol round complexity, thereby increasing the effective bandwidth.}. 
Bandwidth is considered the typical bottleneck for ORAM deployment in client-server applications, 
while memory is the typical bottleneck for the ORAM deployment in trusted execution environments. 
This lack of low-overhead ORAMs, despite considerable efforts from the security community, 
is an undeniable indicator for the need of a fundamentally new approach.

In this paper, we propose a novel approach for developing practical
ORAM protocols. Our key idea is to trade-off performance at the cost 
of quantified statistical privacy. We first formalize the notion 
of a \emph{differentially private ORAM} that provides statistical privacy 
guarantees. As the name suggests, we use the differential privacy framework 
developed by Dwork~\textit{et al.}~\cite{differentialprivacy} with its $(\epsilon,\delta)$-differential 
privacy modification~\cite{epsilondelta}. 
In the current formulation of an ORAM, the output is computationally indistinguishable for any two input sequences.
In a differentially private ORAM, we characterize the effect of a small 
change in the ORAM input to the change in the probability distribution at the output.

This formalization of a differentially private ORAM subsumes the current notion of ORAM security viz., $\epsilon = 0$ leads to the currently accepted ORAM security definition in Section~\ref{sec:statisticaloram}. Yet such a formalization opens up a large underlying design space currently not considered by the community. We also present \ourprotocol{}, a tunable family of ORAM schemes allowing variable bandwidth overheads, 
system security and outsourcing ratios while providing quantified privacy guarantees of differentially private ORAMs. \ourprotocol{} is not a silver bullet for all applications (see Section~\ref{sec:DPPIR} for enabling requirements). But we hope that this first step in the direction of statistically private ORAMs opens the door for the research community to build more efficient ORAM protocols. In Section~\ref{sec:DPPIR}, we demonstrate an application where DP-ORAM provides a promising solution to the problem of private information retrieval.

\subsection{Our Contributions}
\ourprotocol{} introduces a number of paradigm shifts in the design of ORAM protocols while 
building on the prevailing ideas of contemporary ORAM constructions. Our main contributions are: 

\textbf{Formalizing \emph{differentially private ORAMs}: }We formalize the notion of a 
\emph{differentially private ORAM}, which to the extent of our knowledge is 
the first of its kind. A differentially private ORAM bounds the information 
leakage from memory access patterns of an ORAM protocol. 
For details, refer to Section~\ref{sec:statisticaloram}.

\textbf{Tunable protocol family: }We propose a tunable family of ORAM protocols called \ourprotocol{}. These schemes can be tailored as per the needs and 
constraints of the underlying application to achieve a desirable trade-off between security, bandwidth and local storage. This serves as a key enabler for practical deployment 
and is discussed in more detail in Section~\ref{sec:systems}.

\textbf{Security and Utility: }We analyze and provide theoretical guarantees for 
the security offered by \ourprotocol{} schemes in the proposed differentially private 
ORAM framework. The proofs are general and will be useful for analyzing 
the security of alternative statistically private ORAM schemes in the future. 
We also theoretically analyze the utility benefits of using statistical privacy. 
These results are supported by extensive experiments using a complete implementation 
of \ourprotocol{}. The central results of this paper are summarized below (for details, refer to Section~\ref{sec:analysis}). 
\vspace{-6pt} 
\begin{itemize}
\itemsep0em
\item We prove that the family of \ourprotocol{} protocols 
described in Section~\ref{sec:newmodel} satisfy the $(\epsilon, \delta)$-differential 
privacy guarantees and give the relation between $\epsilon, \delta$ and the model parameters.
\item We concretely show the benefits of using differential privacy i.e., we demonstrate 
how a larger value of $\epsilon$ helps reduce the protocol overheads, thereby showing 
an explicit security-performance trade-off.
\end{itemize}
\vspace{-6pt}

\textbf{Practical Impact and Applications:} We experimentally investigate the 
impact of DP-ORAM in the following contexts:
\vspace{-6pt} 
\begin{itemize}
\itemsep0em
\item To reduce local storage requirements to run ORAM protocols 
in trusted hardware. Trusted execution environments such as the first 
generation Intel Skylake SGX processors have stringent memory constraints, 
with available memory for implementing programs (including the ORAM overhead) as low as 90MB~\cite{ohrimenko2016oblivious}. 
For reasonable values of the privacy budget $\epsilon$, \ourprotocol{} reduces 
local storage by more than $2\times$, thereby enhancing compatibility with Intel SGX (Section~\ref{subsec:usability}).
\item To reduce the bandwidth in embedded computing and IoT applications, 
where devices have limited available bandwidth. Depending on the system 
parameters chosen, DP-ORAM can reduce the bandwidth overhead by $2\times$-$10\times$ at 
the cost of statistical security and higher local storage. 
\item We also demonstrate how statistical ORAMs in conjunction with trusted 
hardware can be used to perform differentially private PIR queries~\cite{goldbergDPPIR}. 
We justify the trade-off between enhanced performance at the cost of 
quantified statistical security for PIR protocols in Section~\ref{sec:DPPIR}.
\end{itemize}
\vspace{-6pt}

\ourprotocol{} enables novel design points in developing ORAM protocols by 
leveraging the benefits of statistical privacy. It also supports design 
points with order of magnitude performance improvements over state-of-the-art 
protocols (at the cost of a quantified loss in security). 
Finally, \textit{\ourprotocol{} does not assume any server-side computation} 
and requires practical amounts of client-side storage (depending on the 
parameters chosen). It is also extremely simple to implement at both the client 
and the server side.


\section{Differentially Private ORAM}\label{sec:statisticaloram}

The notion of statistical privacy has been around in security/privacy applications~\cite{bellare1993efficient, liang2009information, mittal2012information} yet it has never been previously explored in the context of ORAMs. We believe formulating such a framework would greatly expand the ability of the research community to develop novel ORAM protocols with low-bandwidth and low-client overhead, serving as an enabler for real-world deployment of this technology.

Formally, an ORAM is defined as a protocol (possibly randomized) which takes an input access sequence $\mathbf{a}$ as given below,
\begin{equation}\label{eqn:y}
\mathbf{a} = {\tt ((op_{M},addr_{M},data_{M}),. . . , (op_1,addr_1,data_1)) }
\end{equation}
and outputs a resulting output sequence denoted by $\mathbf{o} = {\tt ORAM(\mathbf{a})}$. Here, $M$ is the length of the access sequence, ${\tt op_i}$ denotes whether the ${\tt i}^{\text{{\footnotesize th}}}$ operation is a read or a write, ${\tt addr_i}$ denotes the address for that access, and ${\tt data_i}$ denotes the data (if ${\tt op_i}$ is a write). Denoting by $|\mathbf{a}|$ the length of the access sequence $\mathbf{a}$, the currently accepted security definition for ORAM security can be summarized as follows~\cite{pathoram}:


\begin{defn}
\textbf{(Currently accepted ORAM Security): } \textit{Let $\mathbf{a}$ as given in Eq.~\ref{eqn:y}, denote an input access sequence. Let $\mathbf{o} = {\tt ORAM(\mathbf{a})}$ be the resulting randomized data request sequence of an ORAM algorithm. The ORAM guarantees that for any two sequences $\mathbf{a}$ and $\mathbf{a}'$, the resulting access patterns ${\tt ORAM(\mathbf{a})}$ and ${\tt ORAM(\mathbf{a}')}$ are computationally indistinguishable if $|\mathbf{a}| = |\mathbf{a}'|$, and also that for any sequence $\mathbf{a}$ the data returned to the client by ORAM is consistent with $\mathbf{a}$ (i.e the ORAM behaves like a valid RAM) with high probability.}
\end{defn}

This framework for ORAMs is constructed with complete security at its core~\cite{pathoram,ringoram,gentryoramSC,burstoram} and there is no natural way to extend this to incorporate a statistical privacy notion. Hence, we introduce and formalize a statistically private ORAM viz., \emph{differentially private ORAM (DP-ORAM)}. 

\subsection{Formalizing DP-ORAM}

The intuition behind a DP-ORAM is that given any two input sequences that differ in a single access, the distributions of their output sequences should be ``close''. In other words, similar access sequences lead to similar distributions. We formally define it as follows:

\begin{defn}
\textbf{Differentially Private ORAM: } \textit{Let $\mathbf{a}$, as defined in Eq.~\ref{eqn:y}, denote the input to an ORAM. Let ${\mathbf{o} = \tt ORAM(\mathbf{a})}$ be the resulting randomized data request sequence of an ORAM algorithm. We say that an ORAM algorithm is $(\epsilon,\delta)$-differentially private if for all input access sequences $\mathbf{a_1}$ and $\mathbf{a_2}$, which differ in at most one access, the following condition is satisfied by the ORAM,
\begin{equation}\label{eqn:DPoram}
Pr[{\tt ORAM(\mathbf{a_1})} \in S] \leq e^{\epsilon} Pr[{\tt ORAM(\mathbf{a_2})} \in S] + \delta 
\end{equation}
where $e$ is the base of the natural logarithm and $S$ is any set of output sequences of the ORAM.}
\end{defn}

First we note that we lose no generality by using this definition: it can capture the existing computational ORAM security paradigm using $\epsilon = 0$ and negligible $\delta$. The formalism also does not make any assumption about the size of the output sequences in $S$. If the input to the ORAM is changed by a single access tuple ${\tt (op_{i},addr_{i},data_{i})}$, the output distribution does not change significantly. 
Given two sequences $\mathbf{a_1}$ and $\mathbf{a_2}$, the two distributions generated (the red and the blue) are close to each other in the differential privacy sense.

Differential privacy provides two important composability properties~\cite{differentialprivacy} viz, ``composition'' and ``group privacy''. The former (Theorem~\ref{thm:composability}) refers to the degradation of privacy guarantees over multiple invocations of a differentially private mechanism and the latter (Theorem~\ref{thm:groupprivacy}) refers to the privacy guarantees when neighboring databases differ by multiple entries. Together, they give privacy bounds for arbitrary sequences and provide rigorous privacy guarantees over multiple invocations or when access sequences differ in multiple accesses.


%

\begin{theorem}[Composition for DP-ORAM]\label{thm:composability}
\textbf{Invoking an $(\epsilon, \delta)$-differentially private ORAM mechanism $m$ times guarantees $(m\epsilon, m\delta)$-differential privacy.}
\end{theorem}


\begin{theorem}[Group privacy for DP-ORAM]\label{thm:groupprivacy}
\textbf{An $(\epsilon, \delta)$-differentially private ORAM is $(\epsilon', \delta')$-differentially private for access sequences differing by $m$ accesses where $\epsilon' = m \epsilon$ and $\delta' =  m e^{m-1} \delta$. In other words, given two access sequences $\mathbf{a_1}$ and $\mathbf{a_2}$ that differ in $m$ accesses.
\begin{equation}\label{eqn:DPoramgroupprivacy}
Pr[{\tt ORAM(\mathbf{a_1})} \in S] \leq e^{\epsilon'} Pr[{\tt ORAM(\mathbf{a_2})} \in S] + \delta'
\end{equation}}
\end{theorem}

Theorem~\ref{thm:composability} holds even for adaptive queries as long as the randomness used in each mechanism is independent of each other. Together, \textit{Theorem~\ref{thm:composability} and \ref{thm:groupprivacy} allow us to extend differential privacy guarantees to arbitrary access sequences from the guarantees for a single invocation on access sequences that differ by a single access.} It is important to note since privacy guarantees degrade with both the number of invocations and the worst case hamming distance between access sequences, DP-ORAMs are best suited for applications where the input sequences differ 
in a small number of accesses. \textit{We present a case study of such an application - Private Information Retrieval (PIR) - in 
Section~\ref{sec:DPPIR}.}

PIR is a cryptographic primitive for privately accessing data from a public database. 
ORAM schemes can be used in conjunction with trusted hardware to perform 
PIR queries~\cite{williams2008usable, backes2012obliviad}.
We demonstrate the utility of statistical ORAMs by showing how DP-ORAM can be used in 
conjunction with trusted hardware to perform efficient DP-PIR queries~\cite{goldbergDPPIR}. 
This application is well suited to showcase the benefits of using statistical ORAMs as 
\textit{each PIR query corresponds to an access sequence of exactly one element.}

\section{\ourprotocol{} overview}\label{sec:overview}

\begin{table}[t]
\centering
\resizebox{\columnwidth}{!}
{\begin{tabular}{cll|} \toprule
Symbol  & Description \\
\midrule
$N = 2^{L}$      	 	& Number of real data blocks outsourced\\
$0 \leq k \leq L+1$    & Model parameter (to tune bandwidth)\\
$p \in [0, 1]$				& Model parameter (to tune security) \\
$Z$      		    				& Number of blocks in each bucket\\
\bottomrule 
\end{tabular}}
\caption{Notation for \ourprotocol{}}
\label{table:notationrootoram}
\end{table}

In this section, we describe our key design goals and give an overview of the \ourprotocol{} protocol.

\subsection{Design Goals}

\noindent \textbf{Statistically private ORAMs: }We target protocols that offer performance benefits at the cost of statistical privacy which is quantified using the metric of differential privacy.

\noindent \textbf{Tunable ORAM schemes: }Conventional ORAM schemes operate at specific overheads with full privacy but cannot operate at lower overheads. We aim to provide an ORAM architecture that can be tuned to application requirements and can achieve privacy proportional to system resources such as the bandwidth and local storage.

\noindent \textbf{Rigorous Analysis and Efficiency: }We target systems amenable to rigorous security analysis. At the same time, we aim for efficient systems that can be easily implemented on both client and server side.

Finally, the design should use low storage both at the client as well as the server side. \emph{Server side computation is not always practical and hence we do not assume any such capability.} Next we describe the key ideas of \ourprotocol{} protocol. Over the years, several different definitions have been used to quantify ORAM bandwidth overhead. 

We will use the original and straightforward definition of bandwidth as the average number of blocks transferred for one access~\cite{burstoram}.

\begin{defn}\label{def:bandwidth}
\textit{The bandwidth cost of a storage scheme is given by the average number of blocks transferred in order to read or write a single block.}
\end{defn}

\subsection{Approach Overview}
\ourprotocol{} protocol can broadly be split into four components, the storage, the access, the new mapping and the eviction. These are briefly described below. The notation for Root ORAM is illustrated in Table~\ref{table:notationrootoram}. Tree-based ORAMs make a relatively easy proof-of-concept to demonstrate the benefits of DP-ORAMs and hence we construct \ourprotocol{} as a tree-based ORAM.

\begin{figure*}
\center
\includegraphics[width=0.80\linewidth]{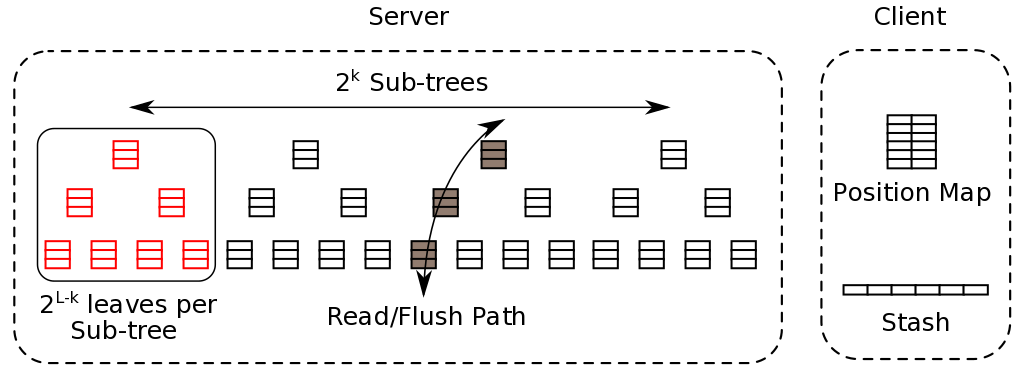}
\vspace{-5pt}
\caption{\textbf{The figure illustrates the client and server side storage. At the server side, there are $2^k$ sub-trees, each having a depth of $L-k$. An individual sub-tree is boxed and shown in red. }}
\label{fig:rootoram}
\vspace{-4pt}
\end{figure*}

\textbf{Storage: }The data to be outsourced is assumed to be split into units called blocks. Blocks are stored at the server-side storage in $2^k$ binary trees, each with a depth of $L-k$. For simplicity of proofs, we call each of these trees as ``sub-trees'' as they can be thought of as sub-trees of a larger virtual tree (cf Fig.~\ref{fig:rootoram}). Each node is a bucket that can hold up to $Z$ data blocks ($Z$ is typically a small constant such as 4 or 5). This is represented in Fig.~\ref{fig:rootoram}. A stash at the client is used to store a small amount of data. Each data block is mapped to a leaf and this mapping is stored recursively in smaller ORAMs.

\textbf{Access: }The main invariant is that any data block is along the path from the associated leaf to the corresponding sub-tree root or is in the stash (as shown in Fig.~\ref{fig:rootoram}). Hence, to access a data block, the client looks up the mapping to find the sub-tree and the associated leaf that the data block is mapped to and then traverses the path from that leaf to the sub-tree root.

\textbf{New Mapping: }The data block is then read or written with the new data and then mapped to a new leaf. It is important to note that this new mapping is not uniform among the leaves. The flexibility and the choice of this non-uniform distribution is given in Section~\ref{sec:newmodel}.

The intuition behind using a non-uniform distribution is that it provides performance benefits such as improving stash storage (refer to Theorem~\ref{thm:DPhelps}). At the same time, we theoretically quantify the security impact of non-uniform distributions using the framework of differential privacy (refer to Theorem~\ref{thm:epsilon}).

\textbf{Eviction: }Finally, new randomized encryptions are generated and all the data (including some blocks from the stash) are written to the accessed path with blocks being pushed as further down the path as possible (towards the leaf). \ourprotocol{} also uses the recursion technique developed previously~\cite{tree_based_orams, SSSoram, pathoram} to store the mapping in smaller ORAMs.

\subsection{Comparison with Path ORAM~\cite{pathoram}}
\ourprotocol{} is a generalization of the Path ORAM protocol~\cite{pathoram}, yet there are critical differences between the two protocols. In this subsection, we highlight some of the critical differences between the two papers.

\textbf{Differentially Private ORAM: }\ourprotocol{} introduces a new metric to quantify ORAM security, which extends the current formalism to include the notion of a statistically private ORAM. We bound the privacy offered by the \ourprotocol{} (as well as Path ORAM) using this metric.

\textbf{Tunable Statistical ORAM: }Path ORAM incurs a fixed bandwidth cost that cannot be tuned. Thus, applications that cannot accommodate high bandwidth costs are unable to achieve access pattern security. \ourprotocol{} on the other hand is tunable and applications with limited bandwidth can achieve security proportional to their resources. 

\textbf{Multi-dimensional design space: }We demonstrate the feasibility of new design points by showing a multi-dimensional trade-off between bandwidth, security and client storage. We support a range of operating conditions by tuning the protocol parameters and demonstrate the trade-off between resource overheads and statistical privacy both theoretically and experimentally.

Note that Path ORAM is an instantiation of \ourprotocol{} for $k = 0$. For more details, see the remark at the end of Section~\ref{sec:systems}.
\section{\ourprotocol{} details}\label{sec:newmodel}

In this section, we provide the details of \ourprotocol{}. Basic notation is given in Table~\ref{table:notationrootoram}. $B$ denotes size of each block in bits, $P(x)$ denotes path from leaf $x$ to the sub-tree root, $P(x,i)$ the node at level $i$ in $P(x)$ and $x := {\tt position[a]}$ indicates data block ${\tt a}$ is currently mapped to leaf $x$.

\subsection{Server Storage}
\textbf{Server Storage: }The server stores data in the form of $2^k$ binary trees as shown in Fig.~\ref{fig:rootoram}. Each node of the tree is a bucket containing multiple data blocks, real or dummy (a dummy block is a randomized encryption of 0). For the simplicity of analysis, we consider the roots of these sub-trees to be at level $k$ and subsequent levels $k+1, \hdots L$ where $L$ would correspond to the leaves of each sub-tree.

\textbf{Bucket structure: }Each node is a bucket consisting of $Z$ blocks, each block can either be real or dummy (encryptions of $0$). 

\textbf{Path structure: }The leaves are numbered in the set $\{0,1,. . ., 2^{L} - 1\}$. $P(x)$ denotes the path (set of buckets along the way) from leaf $x$ to the sub-tree root and $P(x,i)$ denotes the bucket in $P(x)$ at level $i$. It is important to emphasize here that the path length in \ourprotocol{} is $(L+1 - k)$ blocks compared to the $L+1$ blocks in Path ORAM.

\textbf{Dummy blocks and randomized encryption: }We use the standard padding technique (fill buckets with dummy blocks when needed) along with randomized encryption to ensure indistinguishability of real and dummy blocks.

\subsection{Invariants of the scheme}\label{subsec:updatemapping}
\textbf{Main Invariant: }The main invariant in \ourprotocol{} is that each \textit{real} data block ${\tt a}$ is mapped to a leaf $x := {\tt position[a]} , x \in \{ 0,1,2,. . . , 2^L-1\}$ and at any point in the execution of the ORAM, the real block will be somewhere in a bucket $\in P(x)$ or in the local Stash. This path is from the root of a sub-tree to the leaf $x$ and consists of $L-k+1$ buckets. It is also important to note that the invariant does not say that the mapping of each data block is uniform over the set of leaves, as shall be clarified by the second invariant.

\begin{figure}
\centering
\includegraphics[width=0.9\linewidth]{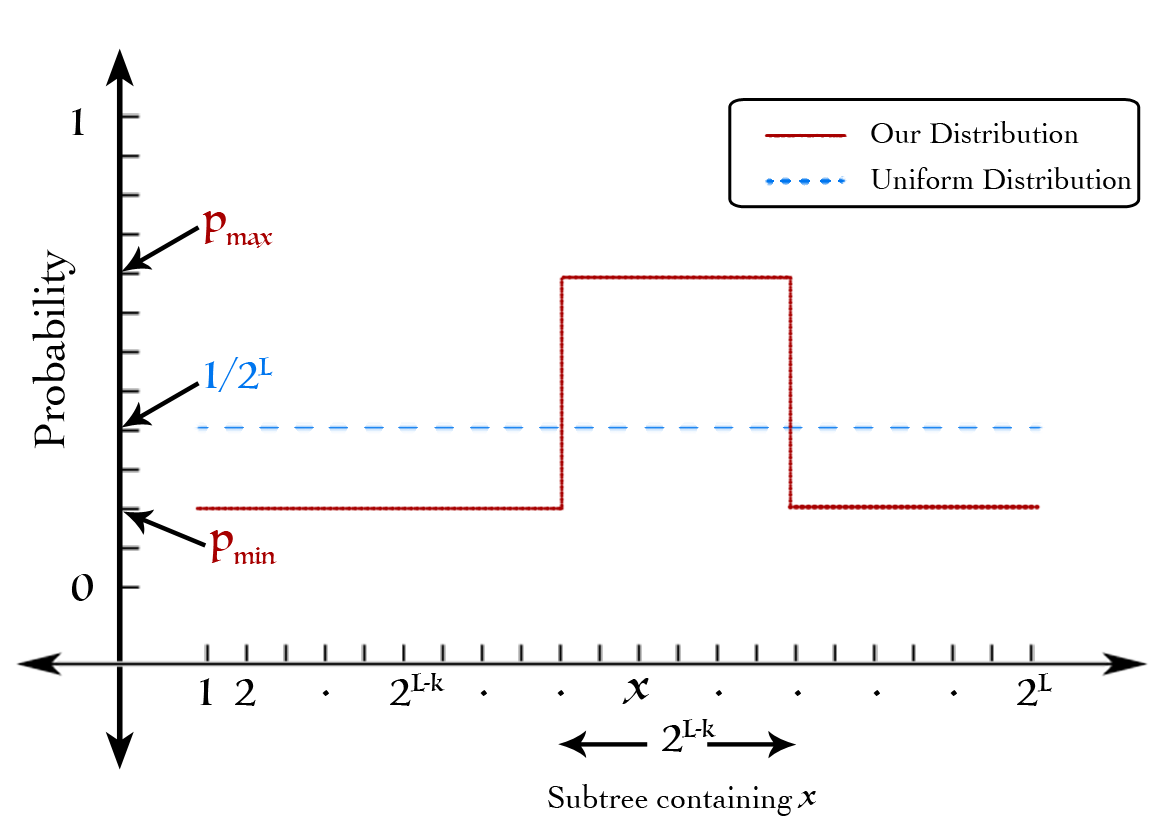}
\vspace{-5pt}
\caption{\textbf{New mapping of a block is more likely among the current sub-tree than any other sub-tree (red). Alternative distributions turn out to be sub-optimal for \ourprotocol{}.}}
\label{fig:distributionD}
\end{figure}

\textbf{Secondary Invariant: }We maintain the secondary invariant that after each access to a data block, its mapping changes according to a leaf dependent non-uniform distribution $D$ (i.e., its new mapping is randomly sampled from this distribution $D$). There is tremendous flexibility in choosing this distribution; for our purposes, we consider a distribution in which a data block is more likely to be remapped to another leaf in the same sub-tree than to another sub-tree leaf. This distribution $D$ is formally given by Eq.~\ref{eqn:distributionD} and shown graphically in Fig.~\ref{fig:distributionD}.
\begin{equation}\label{eqn:distributionD}
P_{z,x} = p_{{\tiny \mbox{min}}} + (p_{{\tiny \mbox{max}}} - p_{{\tiny \mbox{min}}}) \delta_{r_z r_x}
\end{equation}
Where $P_{z,x}$ is the probability that the new mapping is leaf $z$ given the previous mapping was leaf $x$, $r_x$ denotes the root of the sub-tree of leaf $x$, $\delta_{ij}$ is the Kronecker delta defined as
\begin{equation*}
\delta_{ij} = 
\begin{cases}
0 	& \text{if $i \neq j$}\\
1	& \text{if $i=j$}
\end{cases}
\end{equation*} and $p_{{\tiny \mbox{max}}}$ and $p_{{\tiny \mbox{min}}}$ are functions of the model parameter $p$ and are given by: 
\begin{equation}\label{eqn:probabs}
\begin{aligned}
p_{{\tiny \mbox{max}}} &= \frac{1+ (2^k - 1)p}{N} \\
p_{{\tiny \mbox{min}}} &= \frac{1- (1-\delta_{k0}) p}{N}
\end{aligned}
\end{equation}

The reason behind using a non-uniform distribution is that it gives performance benefits such as lower stash usage, captured theoretically in Theorem~\ref{thm:DPhelps}. This particular choice of $D$ happens to be ideal for \ourprotocol{} as can be seen by the analysis from Section~\ref{sec:analysis}. Theorem~\ref{thm:epsilon} gives the relation between the model parameter $p$ and the desired level of privacy (given by $\epsilon$). In practice, the acceptable privacy budget $\epsilon$ would decide the parameter $p$ used in the model. We refer the reader to Section~\ref{subsec:choose} for details of choosing the parameters. 

\subsection{Client Storage}
\textbf{Position Map: }The client side stores a position map which maps real data blocks to leaves of the server tree. This position map is stored recursively using smaller ORAMs. The recursion technique~\cite{tree_based_orams, SSSoram, pathoram} aims to recursively store the ORAM position maps into subsequently smaller ORAMs. The final ORAM position map is stored locally.

\textbf{Stash: }The client maintains a local stash, which is a small amount of storage locally at the client which is used to store overflown data blocks locally\footnote{The stash can be stored on the server at the cost of an increase in the bandwidth whereas in our approach we provide a way to reduce the stash without impacting the system bandwidth and hence store the stash locally.}.\vspace{5mm}

\subsection{Protocol Details}
The main functions of the protocol are Access and updateMapping. 
In the former, we read blocks along a path of a sub-tree, try to write blocks back to the same path (with new encryptions) and if there is insufficient storage, the excess data blocks are stored locally in the Stash. The latter function generates a distribution where a data block is more likely to be remapped to another leaf in the same sub-tree than to another sub-tree leaf. We use sub-tree$_a$ to denote the sub-tree the data block $a$ is currently mapped to, $\cup$ to denote union and $\backslash$ to denote removal from stash.

\vspace{2mm}
\noindent\framebox[.486\textwidth]{%
\begin{minipage}{\dimexpr\linewidth-2\fboxsep-2\fboxrule\relax}%
\begin{algorithmic}[1]
\item[  \underline{Access (${\tt op,a,data^*)}$ :}]
	\State $x \leftarrow {\tt position[a]}$
	\State ${\tt position[a]}$ = updateMapping$({\tt a})$
	\State Stash = Stash $\cup \ P(x)$
	\State ${\tt data} \leftarrow$ Read(${\tt a}$) from Stash
	\If{${\tt op = write}$}
		\State Stash = Stash $\backslash ({\tt a, data}) \cup ({\tt a, data^*})$
	\EndIf
	\State flush($x$)
	\State \Return ${\tt data}$
\end{algorithmic}
\end{minipage}
}\vspace{2mm}

The updateMapping function implements the new mapping function as described in Section~\ref{subsec:updatemapping}. The values of the probabilities are as shown in Fig.~\ref{fig:distributionD} (and Eq.~\ref{eqn:probabs}).

\vspace{2mm}
\noindent\framebox[.486\textwidth]{%
\begin{minipage}{\dimexpr\linewidth-2\fboxsep-2\fboxrule\relax}%
\begin{algorithmic}[1]
\item[  \underline{updateMapping (${\tt a)}$ :}]
	\State $x \leftarrow$ UniformReal(0,1)
\If{$ x \leq N \cdot p_{{\tiny \mbox{min}}}$}
	\State \Return Uniform$(0,1,\hdots , 2^L - 1)$
\Else
	\State \Return Uniform$(\rm{sub} \minus \rm{tree}_a)$
\EndIf
\end{algorithmic}
\end{minipage}} \vspace{1mm}

The flush($x$) function is implemented by writing blocks from the stash into the sub-tree, along the path from the associated leaf to the sub-tree root while writing them as low in the sub-tree as possible. The pseudocode for flush($x$) is given below.

\vspace{2mm}
\noindent\framebox[.486\textwidth]{%
\begin{minipage}{\dimexpr\linewidth-2\fboxsep-2\fboxrule\relax}%
\begin{algorithmic}[1]
\item[  \underline{flush ($x$) :}]
\For {$l = L : k$}
	\State $y \leftarrow P(x, l)$
	\State $S' = \{ {\tt (a', d')} \in \mathrm{Stash\ s.t.\ }  P({\tt position[a']}, l) = y \}$
	\State $S' = \min (|S'|, Z)$ blocks from $S'$
	\State Stash = Stash $\backslash$ $S'$
	\State writeToBucket($y, S'$)
\EndFor
\end{algorithmic}
\end{minipage}} 

\section{Theoretical evaluation}\label{sec:analysis}

\vspace{6mm}
\subsection{Notation}\label{subsec:notation}
We begin by developing some notation to present the central results of this paper. We fix $N = 2^L$ to be the total number of outsourced blocks. We denote by $\mathbb{O}_{k,p}^Z$, the \ourprotocol{} protocol with bucket size $Z$ and model parameters $k, p$. We define the sequence of load/store operations by $\textbf{s} = (\textbf{a}, \textbf{x}, \textbf{y})$ where $\textbf{a} = \{ a_i \}_{i=1}^M$ are the logical block addresses loaded/stored, $\textbf{x} = \{ x_i \}_{i=1}^M$ is the sequence of leaf labels seen by the server and $\textbf{y} = \{ y_i \}_{i=1}^M$ is the sequence of new leaf labels.

Let ${\tt st}\left[ \mathbb{O}_{k,p}^Z(\textbf{s}) \right]$ denote the random variable which equals the number of real data blocks in the stash after a sequence of load/store operations $\textbf{s}$, $\mathbb{O}_{k,p}^{\infty}$ denote $\infty$-\ourprotocol{}\footnote{Analogous to the $\infty$-ORAM in~\cite{pathoram}, refer Section~\ref{subsec:performanceresults}.} and ${\tt st}^Z [\mathbb{O}_{k,p}^{\infty} (\textbf{s}) ]$ denote the stash usage after greedy post-processing. The greedy post-processing takes an $\infty$-\ourprotocol{} and reassigns blocks so that each bucket has no more than $Z$ blocks (for details refer to~\cite{pathoram}). The main results of this paper are split into two main categories, Section~\ref{subsec:securityresults} which states and proves the security results and Section~\ref{subsec:performanceresults} which states and proves the performance results.

\subsection{Security Results}\label{subsec:securityresults}
\begin{theorem}[Differentially Private Protocols]\label{thm:epsilon}
\textbf{The \ourprotocol{} protocol with parameters $k,p$ is $(\epsilon, \delta)$-differentially private for the following choice of $\epsilon$ and $\delta$
\begin{equation}
\begin{aligned}
\epsilon &= 2 \log \left( \frac{1 + (2^k - 1) \cdot p}{1- (1-\delta_{k0}) p} \right) \\
\delta &= M \cdot \left( \frac{1+ (2^k - 1) \cdot p}{N} \right)^{M}
\end{aligned}
\end{equation}
where $\delta_{k0}$ is the Kronecker delta, $M$ is the size of the access sequence and $M >$ total stash size.}
\end{theorem}

\textbf{Proof of Theorem~\ref{thm:epsilon}: }Using a conservative security analysis, we prove the bounds on \ourprotocol{} protocols given in Theorem~\ref{thm:epsilon}. The proof is split into two components, viz., the $\epsilon$ bound and the $\delta$ bound. For the $\epsilon$ bound, we first set up the differential privacy framework, then a model to evaluate probabilities of a given input sequence leading to a specific output sequence and finally compute the maximum change that could result from a change in the input. For the $\delta$ bound, we first demonstrate the significance and need for $\delta$ in the security bound and then proceed to conservatively prove the $\delta$ bound.

\vspace{-2em}
\subsubsection{\textbf{The $\epsilon$ bound}: }
\vspace{-1em}

We follow the notation described in Section~\ref{subsec:notation}. We consider two input sequences $\mathbf{a_1}$ and $\mathbf{a_2}$ that differ in only one access, say $a_j$, for some $j \in \{ 1, 2, \hdots , M \}$. We know that the server sees a sequence \textbf{x} given by 
$$
\boldsymbol{\mathrm{x}} = ({\tt position}_M[a_M], \hdots , {\tt position}_1[a_1])
$$
where $x_i := {\tt position}_i[a_i]$ is the position of address $a_i$ for the $i^{{\tiny \mbox{th}}}$ load/store operation, along with the associated path to the root of the sub-tree. Now, we need to compute the ratio of probabilities that ${\tt ORAM}(\mathbf{a_1})$ and ${\tt ORAM}(\mathbf{a_2})$ both lead to the same observed sequence $\mathbf{x}$ at the server. In other words, we compute 
$$
\Pr [{\tt ORAM}(\mathbf{a}) = \mathbf{x}] 
$$
for any set of input sequence and observed leaf sequence $\mathbf{a}$ and $\mathbf{x}$ respectively, of a given fixed size $M$.

We evaluate the above probability by invoking the secondary invariant viz., after each access the mapping of that data block changes randomly according to a fixed distribution $D$ given in Eq.~\ref{eqn:distributionD}. Under this invariant, the probability that a sequence of load/store operations \textbf{a} leads to a particular observed sequence \textbf{x} can be computed according to the rules below. Since the position map of each location changes independently and randomly according to $D$, we can compute the probability that the input sequence \textbf{a} leads to the output sequence \textbf{x} $(\Pr [{\tt ORAM}(\mathbf{a}) = \mathbf{x}])$ by simply multiplying the probabilities of each individual access. This is shown graphically in Table.~\ref{table:probabilitymodel}.
\vspace{-6pt}
\begin{enumerate}
\item If the block is accessed for the first time, its location is random and hence the probability is $1/2^L$.
\item If the block $a_k$ was accessed previously at $a_i$, then the probability is $p_{{\tiny \mbox{max}}}$ or $p_{{\tiny \mbox{min}}}$ depending on whether ${\tt position}_k[a_k]$ and ${\tt position}_i[a_i]$ belong to the same sub-tree.
\item Finally, we multiply all the above probabilities for access 1 to $M$.
\end{enumerate}
\vspace{-6pt}
If at any point during the above enumeration the stash size exceeds $S$, we set the probability to 0. Refer to Section~\ref{attack} for details. The probability that the stash size exceeds $S$ is bounded by Theorem~\ref{thm:stashbounds}.

\begin{table}
\centering
\resizebox{0.48\textwidth}{!}{
\begin{tabular}{ c | c c c c c c c c}
Obs. Seq. & {\color{blue} a'} & b & {\color{blue} a} & c & a & b & d \\ 
Real Seq. &  {\color{red} x} & y & \boxed{x} & z & y & z & {\color{red} x} \\ \bottomrule \toprule
Probability &  $\frac{1}{N}$ & $\frac{1}{N}$ & $p_{{\tiny \mbox{max}}}$ & $\frac{1}{N}$ & $p_{{\tiny \mbox{min}}}$ & $p_{{\tiny \mbox{min}}}$ & $p_{{\tiny \mbox{min}}}$ \\
\end{tabular}}
\caption{\textbf{Demonstration of probabilities given real and observed access patterns \textbf{a, o} respectively. Different symbols for real and observed access patterns are merely for the sake of clarity. Primed symbols are used to denote leaves belonging to the same sub-tree (ex: a, a'). Only the blue symbols affect the probability of the boxed data block. The red elements show the previous and next access of the boxed data block.}}
\label{table:probabilitymodel}
\end{table}

Next, we compute the maximum change in probabilities over two neighboring access sequences $\mathbf{a_1}$ and $\mathbf{a_2}$ that differ in the $i^{{\tiny \mbox{th}}}$ access. Let the logical address accessed in the two sequences be  ${\tt a}$ and ${\tt b}$ respectively i.e., the $i^{{\tiny \mbox{th}}}$ access in $\mathbf{a_1}$ is ${\tt a}$ and in $\mathbf{a_2}$ is ${\tt b}$. Since $\mathbf{a_1}$ and $\mathbf{a_2}$ agree in all other locations, let the previous location of access of block ${\tt a}$ be $l_{pa}$ (leaf $pa$) and the next location be $l_{na}$. Similarly, let $l_{pb}$ denote the previous location of access of ${\tt b}$ and $l_{nb}$ the next location. If any of these 4 do not exist i.e., the symbol was never accessed before or was never accessed afterwards, we define that leaf to be $0$ for the sake of clarity of the equations (if data element ${\tt a}$ was never accessed after the location of access change, then $l_{na} = 0$). Let $l$ be the location of the access of the $i^{{\tiny \mbox{th}}}$ access. Note that $l_{pa}, l_{na}, l_{pb}, l_{nb}, l$ all are specific leaves from \textbf{x} and hence are same for  $\mathbf{a_1}$ and $\mathbf{a_2}$.

It is easy to see that the probabilities can differ in at most 3 places viz., $l$, $l_{na}$ and $l_{nb}$ (probabilities for $l_{pa}$ and $l_{pb}$ depend on the previous access and hence do not change). To make the equations crisp, we define the following extension to the Kronecker delta function, 
\begin{equation*}
\delta_{ij} = 
\begin{cases}
0 	& \text{if $r_i \neq r_j$}\\
1	& \text{if $r_i = r_j$} \\
\frac{1/N - p_{{\tiny \mbox{min}}}}{p_{{\tiny \mbox{max}}} - p_{{\tiny \mbox{min}}}}  & \text{if $j=0$}
\end{cases}
\end{equation*}
where $r_x$ is the root of the sub-tree associated with leaf $x$. This modification of the Kronecker delta is for the simplicity of the equations. The modification ensures that if a symbol is accessed for the first time, then its probability evaluates to $1/N$, as it should.

Now if $\Pr [{\tt ORAM}(\mathbf{a_1}) = \mathbf{x}] > 0$ and $\Pr [{\tt ORAM}(\mathbf{a_2}) = \mathbf{x}] >0 $ i.e., both the ratios are well-defined, we can calculate the ratio of the probabilities as:

\[
\frac{\Pr [{\tt ORAM}(\mathbf{a_1}) = \mathbf{x}]}{\Pr [{\tt ORAM}(\mathbf{a_2}) = \mathbf{x}]} = \frac{P_{l,l_{pa}} \cdot P_{l_{na},l} \cdot P_{l_{nb},l_{pb}}}{P_{l_{na},l_{pa}} \cdot P_{l,l_{pb}} \cdot P_{l_{nb},l}}
\]

After observing that $\frac{1/N}{p_{{\tiny \mbox{max}}}} \geq \frac{p_{{\tiny \mbox{min}}}}{p_{{\tiny \mbox{max}}}}$, we can see that this maximum value of the ratio of probabilities occurs when $l_{na}, l, l_{pa}$ belong to the same sub-tree and $l_{pb}, l_{nb}$ belong to a different sub-tree. In this case, the ratio is given by,
\[
\frac{p_{{\tiny \mbox{max}}} \cdot p_{{\tiny \mbox{max}}} \cdot p_{{\tiny \mbox{max}}}}{p_{{\tiny \mbox{max}}} \cdot p_{{\tiny \mbox{min}}} \cdot p_{{\tiny \mbox{min}}}} = \left( \frac{p_{{\tiny \mbox{max}}}}{p_{{\tiny \mbox{min}}}} \right)^2
\]

Evaluating this in terms of our parameters, $p_{{\tiny \mbox{max}}}$ and $p_{{\tiny \mbox{min}}}$ given by Eq.~\ref{eqn:probabs} and plugging this into the differential privacy equation:
\begin{align*}
  \max_{\substack{\mathbf{a_1,a_2}\\ |\mathbf{a_1 - a_2}| = 1}} \frac{\Pr [{\tt ORAM}(\mathbf{a_1}) = \mathbf{x}]}{\Pr [{\tt ORAM}(\mathbf{a_2}) = \mathbf{x}]}  & \leq \left( \frac{p_{{\tiny \mbox{max}}}}{p_{{\tiny \mbox{min}}}} \right)^2 \\
  & = \left( \frac{1+(2^k - 1)p}{1- (1-\delta_{k0}) p} \right)^2
\end{align*}  
It is important to note that the above equation holds for all observed access sequences \textbf{x}. And hence, we can see that \ourprotocol{} guarantees $\epsilon = 2 \log \left( \frac{1+(2^k - 1)p}{1- (1-\delta_{k0}) p} \right)$. This completes the $\epsilon$ bound.\footnote{Another important point to note here is that the above analysis is a worst case analysis and hence it only depends on two probabilities in the distribution $D$ viz., the largest and the smallest probabilities. The same proof goes through for other probability distributions leading to $\epsilon = 2 \log \left( \frac{p_{{\tiny \mbox{max}}}}{p_{{\tiny \mbox{min}}}} \right)$.}

\vspace{-2em}
\subsubsection{\textbf{The $\delta$ bound}}\label{attack}
\vspace{-1em}

In this subsection, we show the need for $\delta$ in quantifying the security. We demonstrate this necessity using generic tree-based ORAM constructions. We assume that the total stash size is $S$.
For demonstration purpose, we construct a minimal working example. Let: 
\begin{equation}
\begin{aligned}
\mathbf{a} &= \mathit{((r,1,\cdot),(r,1,\cdot),. . . , (r, 1, \cdot))}  \mbox{ and }\\
\mathbf{a}' &= \mathit{((r,1,\cdot),(r,2,\cdot),. . . , (r, S+1, \cdot))}
\end{aligned}
\end{equation}
where $r$ denotes the read operation and $\cdot$ denotes data which is not important for the demonstration. In words, one access sequence consists of $S$ accesses to the same element and the second access sequence consists of $S+1$ different accesses to elements $1,2, . . . ,S+1$. 

It can be seen that the sequence $1,1,. .  .,1$ is a possible output sequence of ${\tt ORAM}(\mathbf{a})$. 
It is not hard to see that the same sequence $1,1,. . .,1$ can never occur as ${\tt ORAM}(\mathbf{a}')$. The reason for this is simply because if there are more than $S+1$ data blocks mapped to the same leaf, the tree ORAM invariant is broken. Hence the $S+1$ accesses to the same location cannot all be different elements. 

To demonstrate this further, we consider a situation where a program is using a tree-based ORAM protocol to hide its access pattern. We also assume that the program has the following traits,
\[
\mbox{Access Pattern} = \begin{cases} 1,1,1,. . .,1 & \mbox{if } \mbox{ Secret = 1} \\ 
																 1,2,3, . . ,S+1 & \mbox{if } \mbox{ Secret = 0} 
																 \end{cases}
\]
If $\mathbf{a}$ is the input access pattern, \textit{and we observe a sequence of $S+1$ or more access made to the same location in ${\tt ORAM}(\mathbf{a})$, we can immediately infer that Secret = 1.} 
It is important to note that the probability of an observed sequence can suddenly jump from a non-zero value to 0 with the change of a single accessed block. We quantify this by the $\delta$ in the $(\epsilon,\delta)$-differential privacy framework for ORAMs. 

We compute the maximum probability for a sequence such that some neighboring sequence (i.e., differing in one access) has zero probability. In particular we choose the following two sequences:
\begin{align*}
\mathbf{a_1} &= (1,2,3, . . . ,M) \\
\mathbf{a_2} &= (1,1,1, . . . , 1)
\end{align*}
If $\Pr [{\tt ORAM}(\mathbf{a_i}) = \mathbf{x}] > 0$ for $i = 1,2$, then we have already shown the $\epsilon$ bound and hence $\delta = 0$ ($\delta=0$ if $M \leq S$). So it remains to find the maximum $\delta$ when one of these is $0$. Let us assume that sequence \textbf{x} has zero probability when the input sequence is $\mathbf{a_1}$ (i.e., $\Pr [{\tt ORAM}(\mathbf{a_1}) = \mathbf{x}] = 0$). In this case, $\delta$ is simply the maximum value of $\Pr [{\tt ORAM}(\mathbf{a_2}) = \mathbf{x}]$. Then, a conservative upper bound on $\delta$ can be found by noting the following: at each location, the associated probability is either $p_{{\tiny \mbox{max}}}$ or $p_{{\tiny \mbox{min}}}$ or $1/N$. Since $p_{{\tiny \mbox{max}}}$ is the largest of these, we can get a upper bound on $\delta$ as
\begin{equation}
\delta \leq p_{{\tiny \mbox{max}}}^{M}
\end{equation}

Finally, to complete the proof, we note that the above $\delta$ bound should hold for each possible output access sequence \textbf{x}. To extend this to all possible subsets $\mathbb{S}$, say the support of $\mathbb{S}$ contains $\mathbf{x_1, x_2 \hdots x_s}$ and $s \leq M$. Hence,  
\begin{align*}
\Pr [{\tt ORAM}(\mathbf{a_1}) \in \mathbb{S}] &= \sum\limits_{k=1}^s \Pr [{\tt ORAM}(\mathbf{a_1}) = \mathbf{x_k}] \\
&\leq \sum\limits_{k=1}^s \left( e^{\epsilon} \cdot \Pr [{\tt ORAM}(\mathbf{a_2}) = \mathbf{x_k}] + \delta \right) \\
&= e^{\epsilon} \cdot \left( \sum\limits_{k=1}^s \Pr [{\tt ORAM}(\mathbf{a_2}) = \mathbf{x_k}]  \right) + s \cdot \delta \\
&\leq e^{\epsilon} \cdot \Pr [{\tt ORAM}(\mathbf{a_2}) \in \mathbb{S}] + M \cdot \delta
\end{align*}
This shows that \ourprotocol{} is $(\epsilon, \delta)$-differentially private for $\delta = M \cdot p_{{\tiny \mbox{max}}}^{M}$. This completes the $\delta$ bound.  $\blacksquare$

\subsection{Performance Results}\label{subsec:performanceresults}

\begin{theorem}[Security-Performance Trade-off]\label{thm:DPhelps}
\textbf{Let $\mathbb{A}_{k,p}^Z$ and $\mathbb{B}_{k,q}^Z$ be two \ourprotocol{} protocols\footnote{The $p$-parameters for the two ORAMs are different because the $p$-parameters are linked to the corresponding security parameters $\epsilon_1, \epsilon_2$ as given by Theorem~\ref{thm:epsilon}.} with privacy parameters $\epsilon_1$ and $\epsilon_2$ respectively, with $\epsilon_1 \geq \epsilon_2$. For any given access sequence $a$, let $R_1$ and $R_2$ denote the random variables for the stash usage after a sequence $\textbf{s}$ of load/store accesses using $\mathbb{A}_{k,p}^Z$ and $\mathbb{B}_{k,q}^Z$ respectively. Then, 
\begin{equation}
\mathbb{E}[R_1]\leq \mathbb{E}[R_2]
\end{equation} 
where the expectation is taken over randomness of $\textbf{x}, \textbf{y}$ in $\textbf{s} = (\textbf{a}, \textbf{x}, \textbf{y})$.}
\end{theorem}

\begin{theorem}[Stash Bounds]\label{thm:stashbounds}
\textbf{The probability that the stash size of the DP-ORAM protocol with parameters $k, Z = 5$ exceeds $R + Z \cdot 2^k$ for $R \geq 1$ is bounded by $14 (0.6002)^R$}
\end{theorem}

\begin{theorem}[Bandwidth]\label{thm:bandwidth}
\textbf{The bandwidth of the \ourprotocol{} protocol with parameters $k,p,Z$ is $2 \times Z(L + 1 - k)$ blocks per real access.}
\end{theorem}

We defer the proofs of Theorem~\ref{thm:DPhelps},~\ref{thm:stashbounds}, and~\ref{thm:bandwidth} to Appendix~\ref{sec:theoremproofs} due to space constraints.

\begin{figure*}[!htp]
\centering
\begin{subfigure}[b]{.46\textwidth}
\centering
\includegraphics[width=0.9\linewidth]{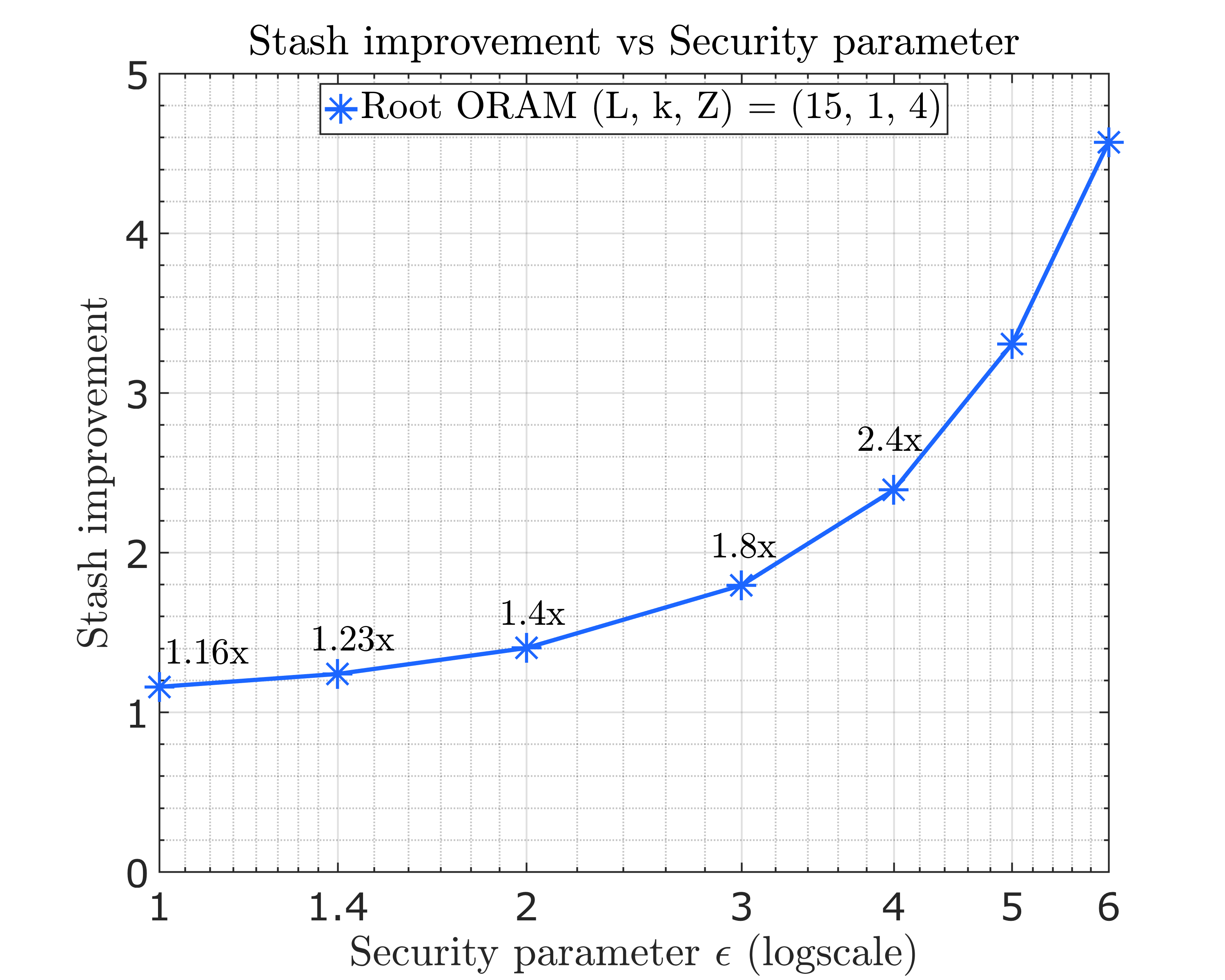}
\caption{}\label{StashSecurity}
\end{subfigure}\qquad
\begin{subfigure}[b]{.46\textwidth}
\centering
\includegraphics[width=0.9\linewidth]{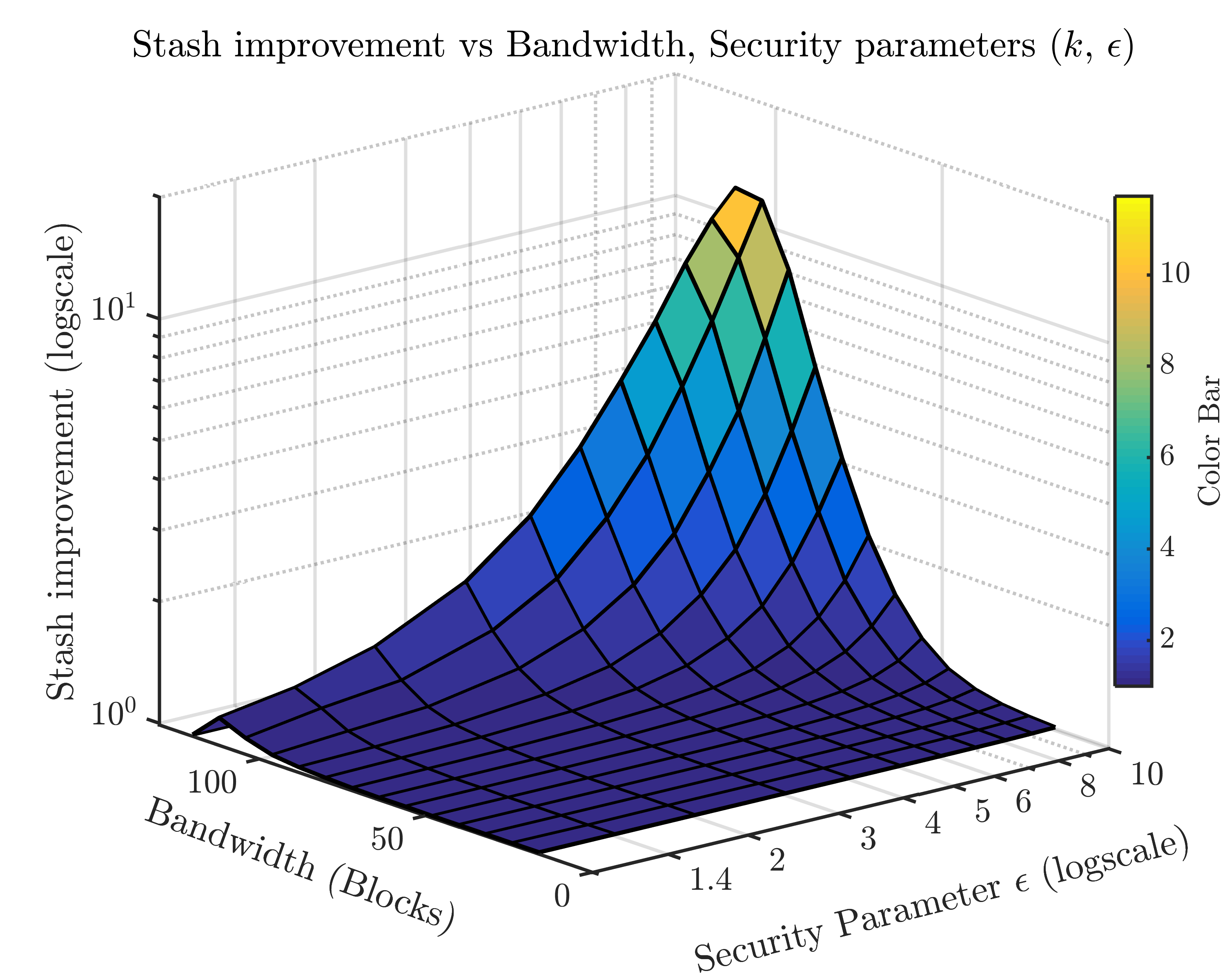}
\caption{}\label{multidimensional}
\end{subfigure}
\vspace{-8pt}
\caption{\textbf{Stash improvement as a function of $\epsilon$ and $k$. Fig.~\ref{StashSecurity} shows the improvement in stash usage compared to a baseline of $\epsilon = 0$ for $(L, k, Z) = (15, 1, 4)$.
Fig.~\ref{multidimensional} shows the security-performance trade-off relative to $\epsilon = 0$ for $(L, Z) = (15, 4)$.}}
\vspace{-8pt}
\end{figure*}

\section{Systems evaluation}\label{sec:systems}

In the previous sections, we have established the design space made possible by formalizing DP-ORAM, a tunable protocol construction, and theoretical security and performance analysis. In this section, we demonstrate the multi-dimensional trade-off between bandwidth, stash storage, and
security using a complete systems implementation of \ourprotocol{}.

\subsection{Details of the implementation}
\ourprotocol{} is built entirely in C++. All experiments were performed on a 1.4 GHz Intel processor with 4GB of RAM. For the Amazon EC2 experiments, remote servers were set-up and latency measurements were performed over a TCP connection for reliable data downloads. For experiments measuring access latency for applications with bandwidth constraints, we cap both the upload and download bandwidth to a value $\gamma$. The values of $\gamma$ used were $\{10, 30, 100, 300, 1000 \}$ KB/s. We used the \emph{trickle} application to constrain the bandwidth at the client machines to these desired values. Finally, we use the worst case linear access pattern for the simulations.

We study the effect of system parameters on the performance of \ourprotocol{}. In particular, we study the inter-dependence between local stash required, bandwidth, and security (given by $\epsilon$). We also study the access latency of \ourprotocol{} protocols in two different settings (1) We measure the access latency over remote Amazon EC2 servers varying the protocol bandwidth parameter $k$ (and consequently the bandwidth itself) (2) We limit the bandwidth at the client end to a specific value $\gamma$ (to emulate constrained bandwidth environments) and measure the access latency of \ourprotocol{} protocols. In light of the recent paper by Bindschaedler \textit{et al.}~\cite{naveedpractical}, we base our experimental evaluation by giving due importance to the constants involved in the overheads of the system.

\subsection{Evaluation results}\label{subsec:evalresults}

\textbf{Bandwidth, Security and Stash Trade-offs: }Fig.~\ref{StashSecurity} shows how statistical privacy reduces stash sizes. Note that increasing values of $\epsilon$ lead to lower stash values, the improvement of which is captured by the $y$-axis of Fig.~\ref{StashSecurity}. While Theorem~\ref{thm:DPhelps} shows that relaxing the security improves performance, Fig.~\ref{StashSecurity} \textit{empirically} shows these performance improvements for concrete values of the security parameter $\epsilon$. 
For instance, \ourprotocol{} provides a $16$\% improvement in stash usage for $\epsilon = 1$, a $40$\% improvement for $\epsilon = 2$ and about $80$\% improvement for $\epsilon = 3$ ($\delta \approx M \cdot  2^{-14M}$ where $M$ is the access sequence length). 
As shown in Appendix~\ref{appendix:DPdefense}, the loss in Shannon entropy of the output sequence is small for moderate values of $\epsilon$. For instance, for $L=20$, an $\epsilon = 3$ results in a loss in entropy of roughly $0.3$ bits and for $\epsilon = 2$ the loss is less than $0.1$ bits (compared to $20$ bits without any security).
Furthermore, as seen in Section~\ref{subsec:MultipleQueries}, in the context of the Private Information Retrieval application the use of anonymous communication channels can further reduce the effective privacy values $\epsilon$ by multiple orders of magnitude.
Similar parameter values for differentially private systems are being increasingly adopted by the research community~\cite{goldbergDPPIR} as well as seen in deployed systems such as RAPPOR~\cite{rappor} ($\epsilon = \ln 3$), Apple Diagnostics~\cite{appleDP, appleDPepsilon} ($\epsilon = 2$ for Health information types, $\epsilon = 4$ for Lookup Hints and Safari crash domain detection and $\epsilon = 8$ for Auto-play intent detection) and US census data release~\cite{UScensusJohn, UScensusDP, UScensusDPharvard} ($\epsilon = 8.9$ for \href{http://onthemap.ces.census.gov/}{OnTheMap} LEHD Origin-Destination Employment Statistics (LODES)). Research works which extensively explore the problem of setting privacy budgets state that the adopted privacy budget values range from $0.01$ to $10$ (refer to Table 1 from Hsu \etal~\cite{hsu2014differential} or Fig. $2$ from Lee \etal~\cite{lee2011much}).


Fig.~\ref{multidimensional} depicts the trade-off between the stash improvement (relative to $\epsilon = 0$), security and bandwidth (parameter $k$) for the \ourprotocol{} protocol. We can see the significant performance gains in the high bandwidth regime. Note that the stash size of the \ourprotocol{} protocol (as in Theorem~\ref{thm:stashbounds}) can be split into two components viz., exponential component (bounded by $Z \cdot 2^k$) and the randomness component (bounded by $14 (0.6002)^R$). The former dominates the latter for small values of bandwidth i.e., large values of $k$ and hence the stash-security trade-off is less significant in those regimes, which agrees with the results in Fig.~\ref{multidimensional}. Fig.~\ref{StashSecurity} and Fig.~\ref{multidimensional} thus capture the effect of varying the security ($\epsilon$) on the performance by the reduction in stash size (compared to a baseline of $\epsilon = 0$) and show that \textit{statistical privacy can be used to improve the performance of ORAM schemes.}

\begin{figure}[!tp]
\centering
\includegraphics[width=\linewidth]{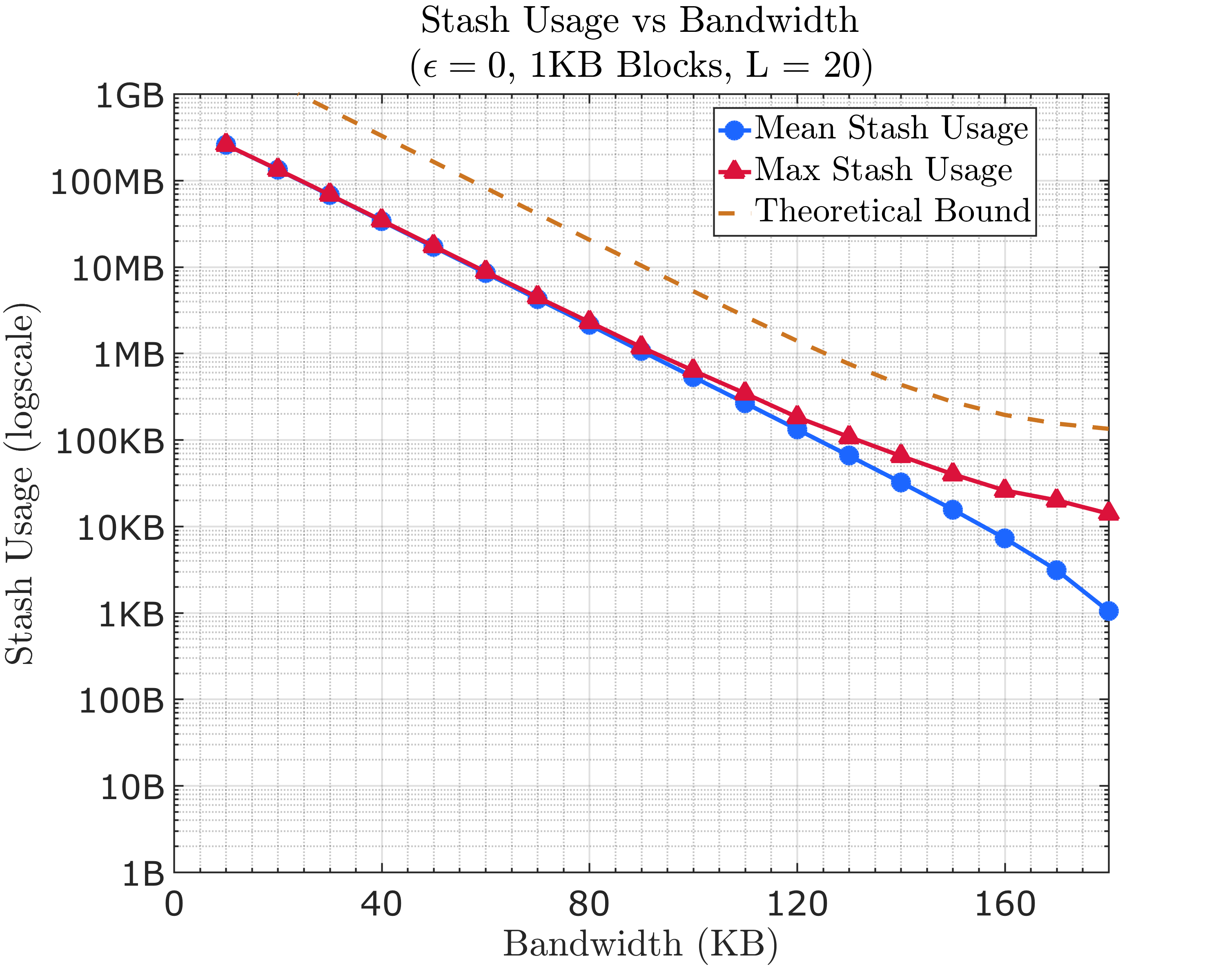}
\caption{\textbf{Absolute values of stash usage for $(L,Z) = (20, 5)$. The theoretical values are plot for a failure probability of $2^{-80}$.}}
\label{fig:stash}
\end{figure}

\textbf{Absolute Stash Values: } Fig.~\ref{fig:stash} shows the absolute values of the stash size (in Bytes) as a function of the bandwidth. The stash roughly grows exponentially with reducing bandwidth, which serves as an experimental validation of Theorem~\ref{thm:stashbounds}. We can see that the required stash values are low enough to be practical in most systems today. For instance, we can achieve a $10\times$ outsourcing ratio at a bandwidth of about 20KB (for 1GB of outsourced data and local storage of 100MB). Similarly, we can achieve $100\times$ outsourcing ratio with a bandwidth of 60KB and an outsourcing ratio of $1000\times$ with a bandwidth of 90KB.

\begin{figure*}[!htp]
\centering
\begin{subfigure}[b]{.47\textwidth}
\centering
\includegraphics[width=\linewidth]{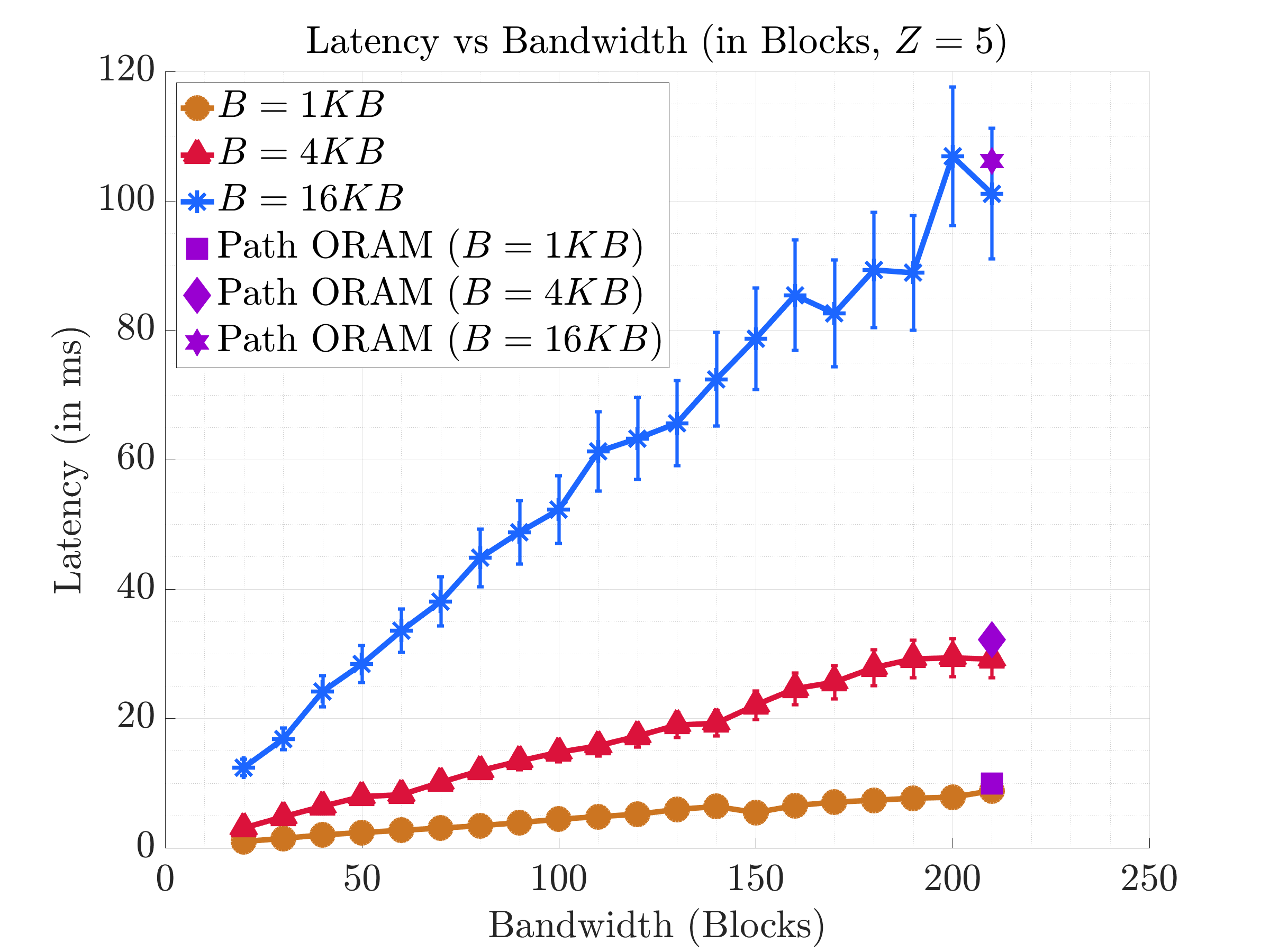}
\caption{}\label{ec2latency}
\end{subfigure}\qquad
\begin{subfigure}[b]{.47\textwidth}
\centering
\includegraphics[width=\linewidth]{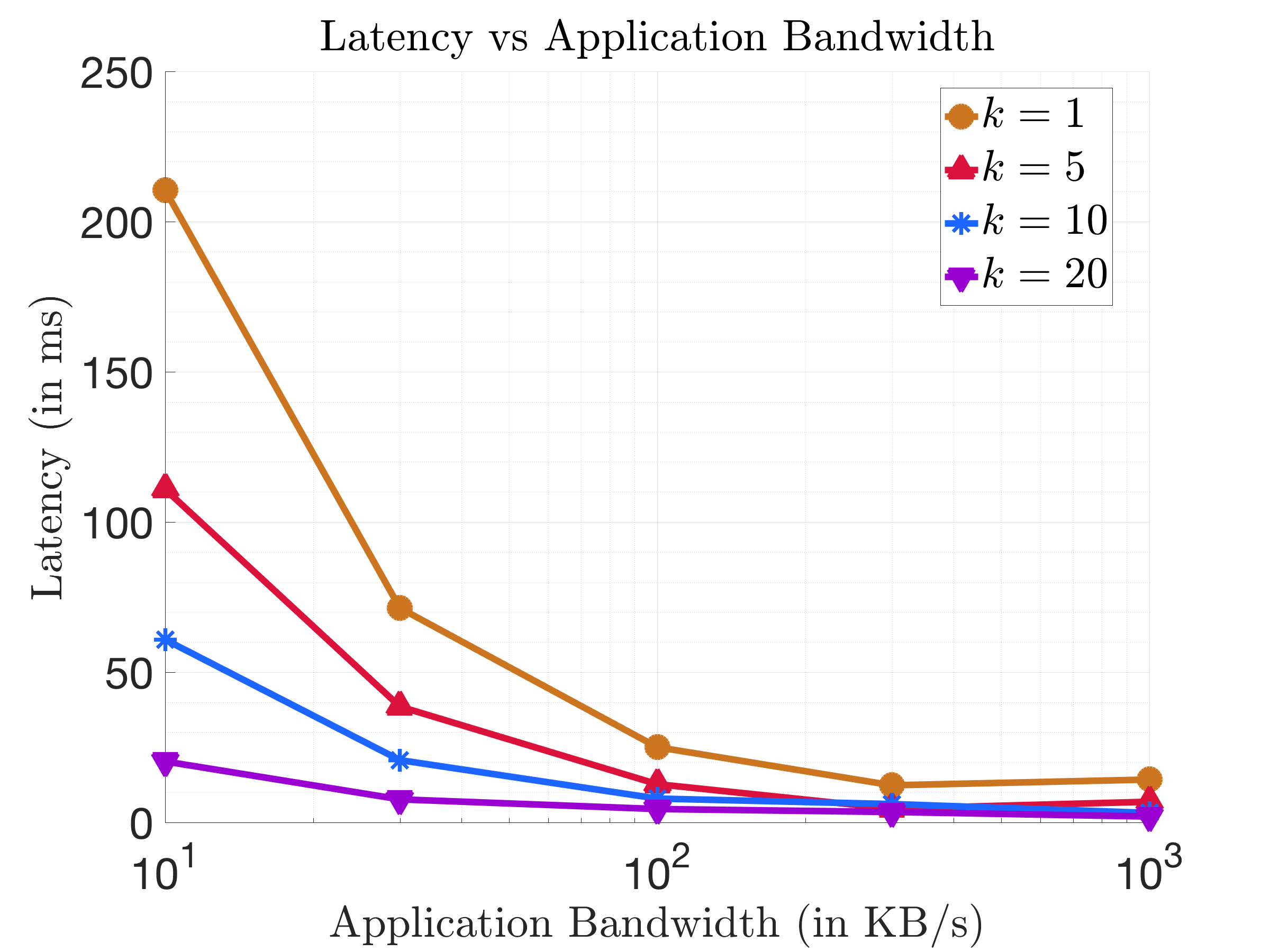}
\caption{}\label{ec2constraint}
\end{subfigure}
\vspace{-10pt}
\caption{\textbf{Real-world implementations over Amazon EC2. Fig.~\ref{ec2latency} shows the \ourprotocol{} latency as a function of the bandwidth for $(L, Z) = (20, 5)$ and different block sizes viz., $1$ KB, $4$ KB and $16$ KB. Fig.~\ref{ec2constraint} shows the latency as a function of the constrained/limited application bandwidth for $4$ KB block sizes and $(L, Z) = (20, 5)$. Note the significant difference between the access latency across different $k$ values for constrained bandwidth applications.}}\label{fig:EC2}
\end{figure*}

\textbf{Real-world implementation: } We compute the latency overhead of a memory accesses as a function of the bandwidth parameter $k$ as well as the constrained application bandwidth $\gamma$.
Fig.~\ref{ec2latency} depicts the access latency as a function of the bandwidth (varying $k$). We can see how \ourprotocol{} provides a spectrum of acceptable bandwidth-latency choices compared to a single design point for Path ORAM. In Fig.~\ref{ec2constraint}, we compare the access latency when the application bandwidth is limited (for a fixed value of $\gamma$). We find the latency as a function of the constrained application bandwidth $\gamma$ (constrained at the client side) for a few different values of $k$. The bandwidth for a given value of $k$ can be computed using Theorem~\ref{thm:bandwidth} as $10*(21 - k)$ blocks. We find that for limited application bandwidth, the system parameters significantly affect the access latency. Hence applications with constrained bandwidths can greatly benefit from using \ourprotocol{}. For instance, in a scenario where the application bandwidth is limited to  10KB/s ($\gamma = 10$KB/s), we can improve the access latency by roughly $10\times$ using \ourprotocol{}.

\subsection{Practical Impact}
\label{subsec:usability}


Next, we consider the significance of local storage and bandwidth improvements offered by \ourprotocol{} in the typical deployment contexts of (1) trusted execution environments and (2) client-server/cloud settings. 

\emph{Local storage:} Trusted Execution Environments, such as enclaves created using Intel SGX-Processors, have severe memory constraints, with total local memory of only 94MB~\cite{ohrimenko2016oblivious}, which is a significant bottleneck for ORAM deployment\footnote{Recent work such as Circuit ORAM~\cite{circuitoram} require constant local memory but increase the protocol round complexity, thereby increasing the effective bandwidth.}. Fig.~\ref{multidimensional} shows 16\% stash improvement for $\epsilon = 1$ values and $1.8 \times$ for $\epsilon = 3$. Hence, for 1KB block size and ORAM parameters $(L, k, Z) = (20,11,5)$, a $2 \times$ improvement would reduce the stash usage from 532KB (Fig.~\ref{fig:stash}) to about 266KB. 
This significantly frees up the limited memory for 
trusted computing operations. 
For larger block sizes such as 256KB (considered in~\cite{naveedpractical}), a $2\times$ stash performance improvement would reduce stash overhead from 133MB to 66.5MB, 
\emph{enabling compatibility with current architectures of trusted processors such as the Intel Skylake}.

Even in the context of smartphone applications, our results indicate that for 1TB of outsourced data, Root ORAM can bring down the local storage overhead (extending Fig.~\ref{fig:stash} results for 1MB block sizes and $(L, k, Z) = (20,11,5)$) from 500MB to less than  250MB (as low as 100MB for higher $\epsilon$).

\emph{Bandwidth:} \ourprotocol{} allows tunable trade-offs between bandwidth, storage, and privacy. In many embedded computing and IoT applications, bandwidth is a significant bottleneck for ORAM deployment. \ourprotocol{} can reduce bandwidth overhead by up to $2\times$-$10\times$ (at the cost of increased local storage and statistical privacy), providing dramatic gains in network access latency as shown in Fig.~\ref{ec2constraint}.\\

\vspace{-2em}
\subsection{Choosing parameters}\label{subsec:choose}
To use \ourprotocol{} as a system, we require a lower bound on the number of accesses $M$ (to bound the worst case $\delta$ leakage). If this is unknown, $M$ is set to $S+1$ (one more than the total stash size). Typical to differentially private systems, a privacy budget is set i.e., an upper bound $\epsilon_{\text{budget}}$ is set for the system use. For the particular application we take into account the worst case hamming distance between access patterns. If this distance is too large, we recommend using $\epsilon_{\text{budget}} = 0$.

Once the privacy budget is set, using the results of Section~\ref{sec:analysis} and Section~\ref{subsec:evalresults}, \ourprotocol{} parameters can be chosen using acceptable values of bandwidth, stash and security parameters ($k , S$ and $\epsilon$). Two of the three parameters can be set to desired values independently viz., two among security parameter $\epsilon$, the bandwidth parameter ($k$) and stash size ($S$) can be chosen independently. The third parameter is determined by the choice of the other two and the optimal trade-off choice would be determined by the specific application requirements. Finally, depending on the application under consideration and the effect of different block sizes on the bandwidth and storage overhead, an optimal block size can be chosen.
For instance, in the application of PIR-Tor~\cite{pirtor}, Tor clients query about 4MB of data from Tor directory servers to retrieve information about Tor relays (refer to Section~\ref{sec:DPPIR} connection between the ORAMs and PIR). 
This can be accomplished by using an ORAM with 4MB block size or a smaller block size ORAM with multiple invocations. The different performance overhead of such choices in system design are quantified in Theorem~\ref{thm:epsilon} and~\ref{thm:stashbounds}, and the resulting security is quantified via composition theorems from Section~\ref{sec:statisticaloram}. 

\textbf{Remark:} It is important to note that when $k = 0$, the storage structure in \ourprotocol{} reduces to a single sub-tree. Hence, the non-uniform distribution in \ourprotocol{} reduces to a uniform distribution over all the leaves. Another sanity check is that both $p_{{\tiny \mbox{max}}}$ and $p_{{\tiny \mbox{min}}}$ equal $1/N$ when $k = 0$. At the same time, since $k=0$, no levels in the tree are cached. \textit{Hence \ourprotocol{} when $k=0$ instantiates exactly into the Path ORAM protocol.}

\section{Applications: Efficient Private Information Retrieval}\label{sec:DPPIR}
In this section, we demonstrate how DP-ORAM in conjunction with trusted hardware 
can be used to perform \emph{differentially private} Private 
Information Retrieval (DP-PIR) queries. The idea of using ORAM in conjunction 
with trusted hardware has been previously explored by the research 
community~\cite{moreno2015privacy, bajaj2011trusteddb, williams2008usable, backes2012obliviad}. 
An important line of research is in developing faster PIR protocols 
using a combination of trusted hardware and ORAM~\cite{williams2008usable, backes2012obliviad}. 

\subsection{Private Information Retrieval (PIR)}
Private Information Retrieval is a cryptographic primitive that provides privacy to a database user. Specifically, the protocol allows the user to hide his/her queries when accessing a public database from the database holder. The critical difference between the PIR and ORAM problem settings is that one assumes a public database (PIR) and the other assumes a private database (ORAM).
In a Differentially Private-PIR scheme (DP-PIR), the PIR privacy guarantees are relaxed and quantified using differential privacy.

\subsection{Differentially Private-PIR schemes (DP-PIR)}
Differentially Private PIR has been proposed by Toledo~\textit{el. al.} in~\cite{goldbergDPPIR}. The definition relies on an indistinguishability game between the adversary and a number of honest users as follows:

\vspace{-6mm}
\subsubsection{DP-PIR indistinguishability game}\label{subsubsec:game}
\vspace{-4mm}

\noindent Among the set of honest users $U$, one is identified by the adversary as the target user $U_t$. The adversary provides the target user $U_t$ two queries $Q_i, Q_j$ and provides all other users a single query $Q_0$. The target user selects one of the two queries and then all users use a PIR system to retrieve records. The adversary observes all the transmitted information including all the information from corrupt servers. The privacy of a DP-PIR protocol is formulated as follows (from Toledo \etal~\cite{goldbergDPPIR}):
\begin{defn}\label{defn:DPPIR}
\textbf{Differentially Private PIR:} \textit{A protocol provides $(\epsilon, \delta)$-private PIR if there are non-negative constants $\epsilon$ and $\delta$, such that for any possible adversary-provided queries $Q_i, Q_j$, and $Q_0$, and for all possible adversarial observations $O$ in the observation space $\Omega$ we have that
\begin{equation}
\forall Q_i, Q_j, \ \mathrm{and} \ Q_0 \qquad \Pr (O | Q_i) \leq e^\epsilon \cdot  \Pr (O | Q_j) + \delta
\end{equation}
}
\end{defn}

The security of DP-PIR schemes translates to the privacy of the underlying queries. Hence, the privacy guarantees of DP-PIR are easier to interpret as they directly relate to the ``program secret'' i.e., the PIR query.

\vspace{-6mm}
\subsubsection{DP-PIR construction from DP-ORAM}\label{subsec:DPPIRProtocol}
\vspace{-4mm}
\noindent To construct a DP-PIR protocol using \ourprotocol{}, we assume the PIR database is on a server with a trusted processor such as Intel SGX~\cite{sgxreference} or 4765 cryptographic co-processor by IBM~\cite{ibmcryptocards}. DP-ORAM based DP-PIR operates on a public database (as required by any PIR application) but is encrypted by the trusted hardware to hide memory accesses. Different users of the DP-PIR application use the same underlying DP-ORAM. 
The DP-ORAM protocol is run within the trusted hardware which also stores the ORAM stash and hence is common across different users and multiple ORAM invocations.
The DP-ORAM block size is set equal to the PIR database block size. To perform a DP-PIR query, a client does the following: 
\begin{itemize}
\itemsep0em
\item \textbf{Step 1 (Initialization):} In the initialization step, the client and the trusted hardware set up an authenticated encrypted channel (AEC) for communication (with or without an anonymous communication channel). The trusted hardware also initializes the ORAM storage structure with the entries of the PIR database. The ORAM is initialized with block size equal to the PIR block size. Other parameters are chosen according to application constraints (refer to Section~\ref{subsec:choose}).
\item \textbf{Step 2 (Send Query): }The client sends his PIR query (some database index $i$) to the trusted hardware through the AEC set up in Step 1 (over an anonymous channel or directly over the network).
\item \textbf{Step 3 (DP-ORAM): }The trusted hardware decrypts the PIR query to get the decrypted index $i$ and initiates a DP-ORAM query using this index.
\item \textbf{Step 4 (Receive Response): }The trusted hardware retrieves the PIR block with index $i$ using the DP-ORAM protocol from the untrusted memory. It sends this block over the AEC to the client. 
\end{itemize}
We show that the above constructed PIR protocol satisfies the guarantees of DP-PIR protocols from Definition~\ref{defn:DPPIR}. More formally,

\begin{theorem}[DP-ORAM $\Rightarrow$ DP-PIR]\label{thm:DPORAMtoPIR}
\textbf{The PIR protocol described above completed using a $(\epsilon, \delta)$-DP-ORAM is $(\epsilon, \delta)$-DP-PIR.}
\end{theorem}

We defer the proof of Theorem~\ref{thm:DPORAMtoPIR} to Appendix~\ref{sec:theoremproofs}.

\subsection{Application Requirements and Multiple Queries:}\label{subsec:MultipleQueries}
\textbf{Application Requirements: }Next we compare the application requirements for various DP-PIR protocols. The 4 DP-PIR protocols from Toledo \etal~\cite{goldbergDPPIR} all rely on the use of multiple servers and 2 of the 4 schemes rely on the use of anonymous communication channels. 
DP-ORAM based DP-PIR described in Section~\ref{subsec:DPPIRProtocol} is a DP-Computational PIR scheme in contrast with the DP-Information-Theoretic PIR schemes in Toledo \etal~\cite{goldbergDPPIR}. Our DP-PIR protocol requires a single server and the use of anonymous channels is optional, though the existence of the latter improves the performance of our proposed protocol as discussed later in this Section. Our protocols require the use of trusted hardware but this results in significant performance improvements as discussed in Section~\ref{subsec:DPPIRcomparison}.

\textbf{Single Queries: }DP-PIR protocols, as formalized in Section~\ref{subsubsec:game}, quantify the privacy for a single PIR query. In Theorem~\ref{thm:DPORAMtoPIR}, we quantify the privacy of our proposed DP-PIR scheme for a single query.
Performance benefits of DP-ORAM directly enhance the performance of the PIR protocol (cf Section~\ref{subsec:DPPIRcomparison}) and showcase the benefits of DP-ORAMs\footnote{DP-PIR is well suited to showcase the benefits of using statistical ORAMs as neighboring sequences in this application directly map to ``program secrets'' and differ by a single access.}.
We further analyze the single query mode of operation into two categories:
\begin{itemize}[noitemsep, topsep=1pt]
\itemsep0em 
\item \textbf{Without anonymous channels (ACs): }Without access to ACs, Theorem~\ref{thm:DPORAMtoPIR} gives the privacy guarantees of our DP-PIR protocol.
\item \textbf{With anonymous channels: }If ACs are available, they can be used to boost the performance of our DP-PIR protocol by leveraging the additional privacy offered by the communication channel. This leads to significant performance benefits which we summarize in the following theorem:
\end{itemize}
\begin{theorem}[DP-PIR with Anonymous Channels]\label{thm:DPPIRAnonymousChannels}
\textbf{The composition of a $(\epsilon, \delta)$-differentially private PIR mechanism with a perfect anonymity system used by $u$ users, for sufficiently large number of users\footnote{Sufficiently large number of users: $u \gg \max \{1,e^{2\epsilon} \}$.}, yields a $(\epsilon', \delta')$-differentially private PIR mechanism for each user where:}
\begin{equation}
\begin{aligned}
\epsilon' &= \frac{e^{2\epsilon}}{u} \\
\delta' &= \min \{1, u \cdot \delta + neg(u)\}
\end{aligned}
\end{equation}
\textbf{where $neg(u)$ is a negligible function\footnote{A negligible function $neg(n)$ is a function $neg \colon \mathbb{N} \to \mathbb{R}$ such that $\forall c \in \mathbb{Z}^+ \quad \exists N_c \in \mathbb{Z}^+$ such that $\forall x \geq N_c$,
$|neg(n)| < n^{-c}$} of $u$.}
\end{theorem}
We defer the proof of Theorem~\ref{thm:DPPIRAnonymousChannels} to Appendix~\ref{sec:theoremproofs}. 
These bounds significantly enhance the privacy values of when using a DP-PIR protocol in composition with a anonymous communication channel. For instance, assuming $u = 10^3$ users use a $(2, 2^{-80})$-DP-PIR, each user is effectively using a $(0.05, 2^{-70})$-DP-PIR protocol\footnote{Ignoring terms negligible in $u$}.


\textbf{Multiple Queries: }
An important consideration in the use of DP-PIR schemes is the effect of multiple queries on the security of the scheme. Multiple invocations of the DP-PIR scheme results in a privacy loss.
We extend Theorem~\ref{thm:composability} to prove Theorem~\ref{thm:DPPIRComposition} that bounds the privacy of DP-PIR schemes under multiple invocations. Consequently, the privacy of multiple DP-PIR invocations can be found by composing Theorem~\ref{thm:DPPIRComposition} with the bounds from Theorem~\ref{thm:DPORAMtoPIR} or Theorem~\ref{thm:DPPIRAnonymousChannels} depending on the availability of anonymous communication channels.
\begin{theorem}[DP-PIR Composition Theorem]\label{thm:DPPIRComposition}
\textbf{$m$ invocations of a $(\epsilon, \delta)$-DP-ORAM based DP-PIR protocol guarantees an overall $(m\epsilon, m\delta)$-DP-PIR protocol.}
\end{theorem}
We defer the proof of Theorem~\ref{thm:DPPIRComposition} to Appendix~\ref{sec:theoremproofs}.

\begin{figure}[t]
\centering
\includegraphics[width=0.95\linewidth]{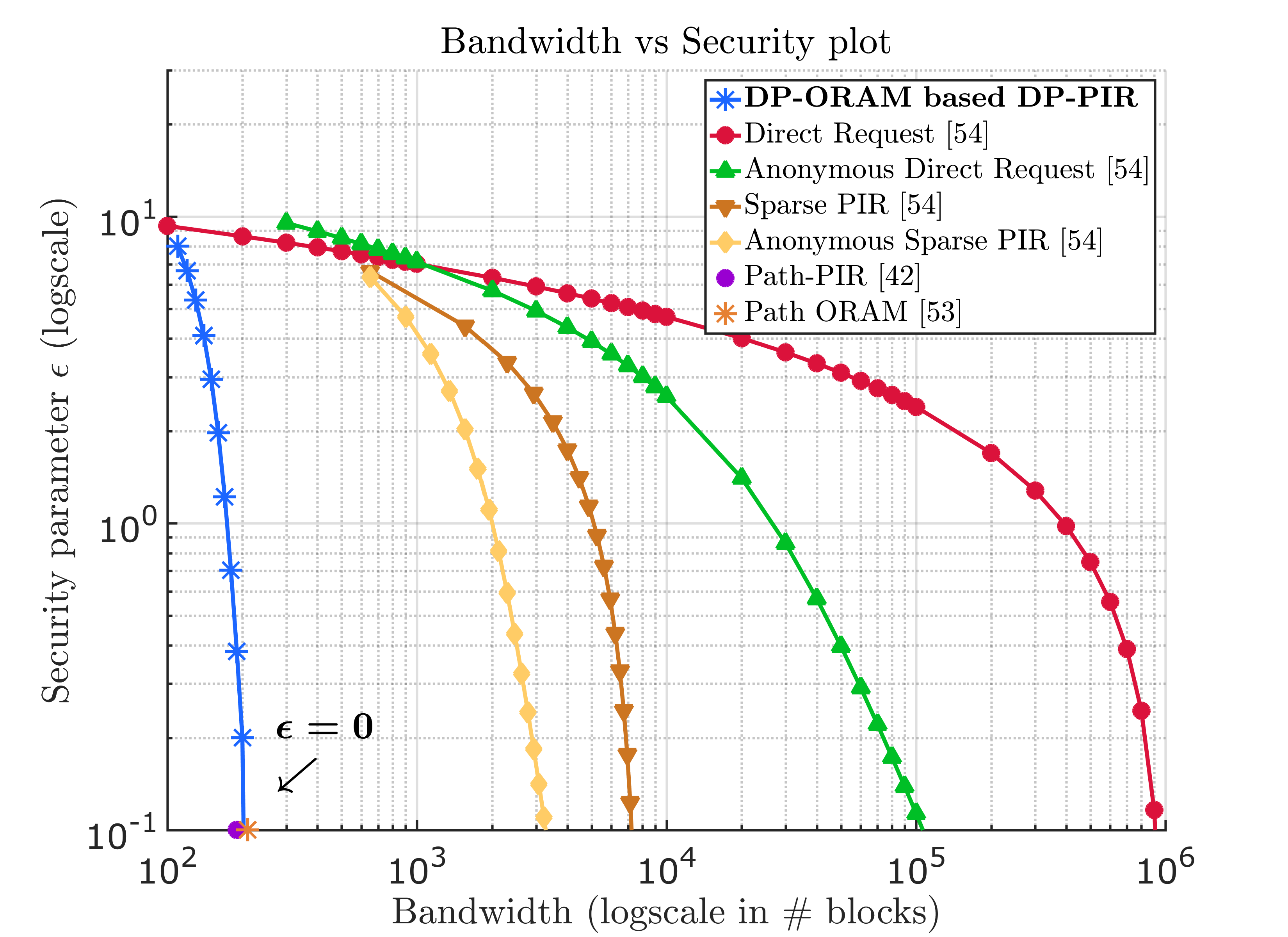}
\vspace{-5pt}
\caption{\textbf{Security-Bandwidth trade-offs for DP-PIR protocols (Toledo \etal~\cite{goldbergDPPIR}, Path-PIR~\cite{mayberry2014efficient}, and Path ORAM~\cite{pathoram}).}}
\label{fig:comparison}
\end{figure}

\subsection{Comparison with Prior Work}\label{subsec:DPPIRcomparison}
Next we compare the performance of our DP-PIR scheme with (1) DP-PIR schemes from~\cite{goldbergDPPIR} (2) Path-PIR construction~\cite{mayberry2014efficient}. We begin by briefly describing the 4 DP-PIR schemes from Toledo \etal~\cite{goldbergDPPIR}:
\begin{itemize}[noitemsep, topsep=1pt]
\itemsep0em
\item \textbf{Direct Requests: }For each real query, the client sends $p-1$ other dummy queries spread across $d$ identical databases. $d_a$ of the databases are assumed to be adversarial.
\item \textbf{Anonymous Direct Requests: }This protocol assumes the use of anonymous communication channels (ACs) and performs the above mentioned Direct request protocol in conjuction with the AC. The increased privacy occurs from the fact that each user sends $p$ requests yet derives privacy among $u\cdot p$ requests (where $u$ is the number of users).
\item \textbf{Sparse-PIR: }This protocol is based on Chor's PIR protocol~\cite{chor}. Instead of generating random vectors for the servers, the client generates biased (hence sparse) random vectors using i.i.d Bernoulli trials with parameter $\theta$. 
\item \textbf{Anonymous Sparse-PIR: }Similar to anonymous direct requests, this protocol is the composition of the Sparse-PIR protocol with an anonymity system.\\
\end{itemize}

We compare our protocols with the exact same set-up as in~\cite{goldbergDPPIR}. Different parameters are set to the following values: (1) Database with $n = 10^6$ blocks (2) Number of databases $d = 10^2$ (3) Number of adversarial databases $d_a = 0.1 \times d = 10$ (this showcases the most optimistic version of the results of~\cite{goldbergDPPIR}). Anonymous direct request has the same parameters as the direct request protocol with an additional assumption of number of users $u = 10^3$. Sparse-PIR and Anonymous Sparse-PIR protocols ignore the communication cost from the client side. This communcication overhead is information theoretically lower bounded by $a\cdot h(\theta)$ where $a$ is the size of vector to be sent and $h(\cdot)$ is the binary entropy function. Since the overhead for encoding the random vectors is linear (and hence very large) in Sparse PIR and Anonymous Sparse PIR protocols, we assume they are based on the 2D variant of Chor's protocol~\cite{chor}. 

\textbf{Bandwidth comparison: }As seen in Fig.~\ref{fig:comparison}, our DP-PIR protocol provides orders of magnitude performance improvements over state-of-the-art DP-PIR protocols from~\cite{goldbergDPPIR}. The performance gains come from the logarithmic overhead of ORAM schemes compared to linear overhead of PIR schemes. Path-PIR does not provide statistical security and hence is seen as a single data-point in Fig.~\ref{fig:comparison}. Path-PIR also achieves logarithmic overhead yet suffers from (1) heavy computation requirements at the client and the server due to the use of underlying homomorphic encryptions (2) large storage overhead due to logarithmic bucket sizes (3) scalability i.e., is better suited for small databases (or large block sizes). 

\textbf{Other comparisons: }As discussed before, our DP-PIR protocol requires the use of trusted hardware but results in significant performance improvements. At the same time, our DP-PIR protocol requires a single server in contrast with multiple servers required for Toledo \etal~\cite{goldbergDPPIR}. It is interesting to note that our DP-ORAM based DP-PIR (described in Section~\ref{subsec:DPPIRProtocol}) is a DP-Computational PIR scheme in contrast with the DP-Information-Theoretic PIR schemes from Toledo \etal~\cite{goldbergDPPIR}. Our protocol as well as protocols from Toledo \etal~\cite{goldbergDPPIR} benefit from the use of anonymous channels. Computational costs for our DP-PIR protocol are $O(\log N)$ whereas they are $O(p), O(d \cdot N \cdot \theta)$ for various schemes in Toledo \etal~\cite{goldbergDPPIR}. On the contrary, additional storage costs for our protocol are given by Theorem~\ref{thm:stashbounds} $(O(\log N) + Z*2^k)$ but are 0 for schemes in Toledo \etal~\cite{goldbergDPPIR}. Finally, we remark that the set-up cost for our protocol includes a \textit{one time} ORAM database initialization\footnote{DP-ORAM initialization is done once and hence can be done using an $\epsilon=0$ DP-ORAM to preserve privacy budget for future queries.} phase (where the PIR database is stored in the ORAM database) which does not exist for other DP-PIR protocols.

\subsection{Other Applications}
Our discussion above focused on the PIR, which itself is a fundamental privacy technology that can enable numerous applications, including PIR-Tor~\cite{pirtor}, PIR for e-commerce~\cite{henry2011practical}, PIR for MIX Nets~\cite{kesdogan2002unobservable}. Benefits of DP-ORAM extend to other applications as well. For instance, Gentry \etal~\cite{gentryoramSC} demonstrate the use of ORAMs as building blocks for secure computation. The benefits of DP-ORAM can be extended to such applications of ORAM protocols for improving performance. In fact, the use of differential privacy to boost the performance of secure computation is already gaining attention in the research community with work by He \etal~\cite{ashwinDPSMC}. Finally, DP-ORAM can be used in systems such as Dropbox and Google Drive to privately retrieve data at low network overheads and local storage.

\section{Related work}\label{sec:relatedwork}


Oblivious RAMs were first formalized in a seminal paper by Goldreich and Ostrovsky~\cite{goldreichoram}. Since then, the research community has made substantial progress in making ORAMs practical~\cite{wsc-bcomp-08,Pinkas2010,gm-paodor-11,GMOT12,pathoram,ringoram,oblivistore}. 
Hierarchical constructions such as~\cite{Pinkas2010,GMOT12,KLO} were proposed building on~\cite{goldreichoram} and tree based ORAM schemes such as~\cite{pathoram,ringoram,gentryoramSC,burstoram,SSSoram,ren15constants,oblivistore} were proposed building on Shi~\textit{et. al.}~\cite{tree_based_orams}. A recent benchmark for ORAMs has been the Path ORAM protocol~\cite{pathoram} that gives theoretical bounds on the local memory usage. Tessaro \textit{et. al.}~\cite{tessarotcc} build on \cite{pathoram} and extend it for multiple clients by level caching in tree based ORAM schemes. \ourprotocol{} generalizes the construction of \cite{pathoram} to provide a tunable framework offering DP-ORAM guarantees. \ourprotocol{} gets around the Goldreich-Ostrovsky lower bound by using (1) Statistical security, which voids the proof of the lower bound~\cite{goldreichoram}
(2) Stash storage that is not a constant (which is what gives the logarithmic GO-lower bound). 
Our work opens up new opportunities for rethinking lower bounds for statistical ORAMs.

Gentry~\textit{et. al.}~\cite{gentryoramSC} have shown the promise of using ORAMs as a building block in developing protocols for Secure Multi-Party Computation. This work is among the first in the line of research using ORAMs as critical component of building other cryptographic primitives. Recently, there has been a number of works using ORAMs for private information retrieval~\cite{moreno2015privacy, bajaj2011trusteddb, williams2008usable,mayberry2014efficient}, for private ad recommendation~\cite{backes2012obliviad} and secure computation and machine learning~\cite{scoram, oblivm, secureml, gentryoramSC}.

Several optimizations have been proposed to reduce the overhead of tree-based ORAMs. Recently, Ring ORAM~\cite{ringoram} reduced the bandwidth using the XOR technique leveraging server-side computation. The XOR technique is orthogonal to the ideas explored in this work and can be extended to \ourprotocol{},
further influencing the protocol design space. Two optimizations for Shi \textit{et. al.}~\cite{tree_based_orams} were proposed by Gentry \textit{et al.}~\cite{gentryoramSC}. First, they reduce the storage overhead by a multiplicative factor and second, they reduce the time complexity of the protocol. They explore the benefits of using a multiple fan-out tree structure instead of a conventional binary tree. ORAM has also been implemented at a chip level in prototypes such as the Ascend architecture~\cite{fletcher2012secure} and the Phantom architecture~\cite{maas2013phantom}. 


Recently, Circuit ORAM~\cite{circuitoram} proposed a novel protocol to reduce the complexity of the eviction protocol in Path ORAM when implementing on a small private memory. This is ideally suited for secure computation environments and is the state-of-the-art protocol when implementing ORAMs in trusted hardware. Though Circuit ORAM works with a constant memory, it increases the protocol complexity which leads to a higher bandwidth usage. Burst ORAM~\cite{burstoram} builds on ObliviStore~\cite{oblivistore} by level caching and optimizing the online bandwidth (formalized in~\cite{boneh2011remote}) for bursty (realistic) access pattern. Onion ORAM~\cite{onionoram} ``breaks'' the ORAM lower bound by leveraging server side computation and additively homomorphic encryptions and achieving constant bandwidth overhead. 

Floram~\cite{floram} is a state-of-the-art ORAM construction which constructs an ORAM protocol in the Distributed ORAM model (DORAM). In the DORAM model, the ORAM memory is split across multiple servers. Whereas in the conventional ORAM setting two logical access sequences of the same length produce indistinguishable physical access sequences, in a DORAM, only the physical access sequences observed by a \textit{single server} are indistinguishable. It is possible to augment our work with Floram to further boost its performance.

In summary, \ourprotocol{} is the first protocol that demonstrates a trade-off between performance and statistical privacy (quantified with differential privacy). The tunable security-bandwidth-outsourcing ratio construction and the formalization of differentially private ORAMs differentiates our work from prior approaches. 
\vspace{-5mm}
\section{Limitations}\label{sec:futurework}
In this work, we enable the design of practical ORAM schemes for applications with 
stringent bandwidth constraints and small local storage. For some applications, it 
might be acceptable to trade-off statistical privacy for better performance and 
\ourprotocol{} demonstrates the first step in this direction by introducing a 
tunable framework that provides differential privacy guarantees. 
Though Theorem~\ref{thm:composability} - \ref{thm:groupprivacy} help us bound the privacy leakage for arbitrary 
access sequences, we acknowledge that \ourprotocol{} is currently better suited for similar access sequences. For example, our approach is ideally suited for applications such as PIR (Section~\ref{sec:DPPIR}). 
The formalization of DP-ORAM opens up a number of research directions such as optimal security-performance trade-offs, rethinking lower bounds for statistically private ORAMs, as well as better performance improvement results. Our work has already inspired other researchers to rethink research ideas at the intersection of differential privacy and conventional cryptography~\cite{mazloom2017differentially,chan2018foundations}.
Finally, we note that the ideas developed in this work are orthogonal yet applicable to more recent works such as Ring ORAM, Onion ORAM, and Burst ORAM~\cite{ringoram, onionoram,burstoram}. Similarly, DP-ORAM constructions for non-tree based ORAMs would be interesting for future work.

\section{Conclusions}\label{sec:conclusion}
To summarize, we introduce and formalize the notion of a differentially private ORAM, which to our knowledge is the first of its kind. We present \ourprotocol{}, a tunable family of ORAM protocols which provide a multi-dimensional trade-off between security, bandwidth and local storage requirements. We evaluate the protocol using theoretical analysis, simulations, and real world implementation on Amazon EC2. 
We analyze the benefits of statistical ORAMs in (1) trusted execution environments and (2) server-client settings and demonstrate how statistical ORAMs can improve the performance of existing ORAMs. Finally, we showcase the utility of \ourprotocol{} via the application of Private Information Retrieval.

\section{Acknowledgments}
We would like to thank the anonymous PETS reviewers for insightful feedback on the paper and the following funding agencies: Army Research Office YIP Award, National Science
Foundation (CNS-1409415, CIF-1617286, CNS-1553437), and Faculty research
awards from Google, Intel, and Cisco.

\bibliographystyle{plain}
\bibliography{./Sameer_bib}
\appendix

\section{Theorem Proofs}\label{sec:theoremproofs}

\textbf{Proof of Theorem~\ref{thm:DPhelps}}: We use two key concepts viz., $\infty$-ORAM and the greedy post-processing algorithm from prior works~\cite{pathoram, circuitoram} in proving the above result. We begin by briefly describing these concepts and then prove an equivalence between a greedily post-processed $\infty$-ORAM and \ourprotocol{} (Lemma~\ref{lem:lemma1} and Lemma~\ref{lem:lemma2}). Finally, we complete the argument by proving the effectiveness of using a non-uniform distribution in reducing the stash usage in $\infty$-ORAM, thereby showing its effectiveness in \ourprotocol{}. Next, we briefly discuss the concepts of $\infty$-ORAM and the greedy post-processing algorithm and refer the reader to~\cite{pathoram} for more details. We follow the notation from Section~\ref{subsec:notation}.

\textbf{$\infty$-ORAM: }This is an imaginary ORAM, used as a mathematical abstraction to facilitate proofs about \ourprotocol{}. The $\infty$-ORAM has all parameters identical to the \ourprotocol{} except it has an infinitely large bucket size ($Z \to \infty$). This allows the $\infty$-ORAM to store as many blocks in a bucket as possible. 

\textbf{Greedy post-processing: }This is an algorithm that post processes the stash and the buckets in an $\infty$-ORAM such that after a sequence of \textbf{s} load/store operations, the distribution of the real blocks over the buckets and stash is exactly the same as that of the \ourprotocol{} after being accessed using \textbf{s}. It is easy to see that the $\infty$-ORAM starts with an empty stash. The greedy post processing algorithm mentioned below processes the $\infty$-ORAM until the tree has no buckets with more than $Z$ blocks. 
\vspace{-5pt}
\begin{itemize}
\itemsep0em
\item Select any block in a bucket that stores more than $Z$ blocks. Suppose that the bucket is at level $h$ and $P$ is the path from the bucket to the root.
\item Find the highest level (closer to the root) $i \leq h$ such that the bucket at level $i$ on path $P$ stores less than $Z$ blocks. If such a bucket exists, move the block to level $i$ else move it to the stash.
\end{itemize}
\vspace{-5pt}

Next, we state Lemma~\ref{lem:lemma1} and Lemma~\ref{lem:lemma2} and omit their proofs due to their similarity with~\cite{pathoram} as well as space constraints.

\begin{lemma}\label{lem:lemma1}
The stash usage in the post processed $\infty$-\ourprotocol{} is the same as \ourprotocol{} protocol with the same parameters.
\begin{equation}\label{eqn:lemma1}
{\tt st}^Z [\mathbb{O}_{k,p}^{\infty} (\textbf{s}) ] = {\tt st}[\mathbb{O}_{k,p}^Z (\textbf{s}) ]
\end{equation}
\end{lemma}
For the sake of analysis, we combine the $2^k$ binary sub-trees by appending a binary tree of depth $k-1$ above the sub-trees. This creates an extended binary tree of height $L$ which contains the original sub-trees at its bottom. We look at the bucket usage over rooted sub-trees of this extended binary tree (rooted sub-tree is a sub-tree which contains the root of the extended tree). We denote by $T$ a generic rooted sub-tree. We use $n(T)$ to denote the total number of buckets in $T$ and ${\tt u}^T(\mathbb{O}^{\infty}_{k,p} [s])$ for the number of real blocks in $T$ for an $\infty$-\ourprotocol{} after a sequence $s$ operations. 
\begin{lemma}\label{lem:lemma2}
The stash usage ${\tt st}^Z [\mathbb{A}_{k,p}^{\infty} (\textbf{s}) ]$ in post-processed $\infty$-\ourprotocol{} is $> R$ if and only if there exists a sub-tree $T$ in $\infty$-\ourprotocol{} such that ${\tt u}^T \left( \mathbb{A}^{\infty}_{k,p} [s] \right) > n(T)\cdot Z + R$
\end{lemma}
Let $\mathbb{A}_{k,p}^Z$ and $\mathbb{B}_{k,q}^Z$ be two \ourprotocol{} protocols with security parameters $\epsilon_1$ and $\epsilon_2$ respectively. 


Suppose $S$ denotes the set of leaves of the extended binary tree and $S’$ the set of leaves of the currently mapped sub-tree. The probability distribution functions for the updateMapping function in $\mathbb{A}_{k,p}^Z$ and $\mathbb{B}_{k,q}^Z$ differ only in the following way: some probability mass $m$ (for some $m \geq 0$) moves from leaves $S-S’$ to $S’$.

Thus, with probability mass ($1-m$), the randomized mapping for both protocol $\mathbb{A}_{k,p}^Z$ and protocol $\mathbb{B}_{k,q}^Z$ behave identically. However, with probability mass $m$, the data block will be mapped to a leaf in $S’$ in protocol $\mathbb{A}_{k,p}^Z$ but to a leaf in $S-S’$ in protocol $\mathbb{B}_{k,q}^Z$. Hence, in the $\infty$-ORAM, the data block will be placed on a level less than $k-1$ (higher up in the tree) in $\mathbb{B}_{k,q}^{\infty}$ whereas in $\mathbb{A}_{k,q}^{\infty}$ it will be placed the same sub-tree i.e., on a level greater than $k-1$ (lower down in the tree). Hence for any subtler $T$, if the data block in $\mathbb{A}_{k,q}^{\infty}$ was placed in a bucket in $T$, then so will the data block in $\mathbb{B}_{k,q}^{\infty}$. Hence,

$$
{\tt u}^T(\mathbb{A}^{\infty}_{k,p} [s]) \leq {\tt u}^T(\mathbb{B}^{\infty}_{k,q} [s])
$$
Hence, for any given sub-tree $T$, we have:
\begin{align*}
\left( {\tt u}^T(\mathbb{A}^{\infty}_{k,p} [s]) > n(T)\cdot Z + R \right) \\
\Rightarrow  \left( {\tt u}^T(\mathbb{B}^{\infty}_{k,q} [s]) > n(T)\cdot Z + R \right)
\end{align*}
Using the above condition over all rooted sub-trees $T$, we have
\begin{align*}
\Pr &\left[ \exists T \left( {\tt u}^T(\mathbb{A}^{\infty}_{k,p} [s]) > n(T)\cdot Z + R \right)  \right]  \\
&\leq \Pr \left[ \exists T \left( {\tt u}^T(\mathbb{B}^{\infty}_{k,q} [s]) > n(T)\cdot Z + R \right)  \right]
\end{align*}
Hence, 
\begin{align*}
\Pr \left[{\tt st} [\mathbb{A}_{k,p}^Z (\textbf{s}) ]\right. &> \left. R \right]  = \Pr \left[{\tt st}^Z [\mathbb{A}_{k,p}^{\infty} (\textbf{s}) ] > R \right]  \\
&= \Pr \left[ \exists T \left( {\tt u}^T(\mathbb{A}^{\infty}_{k,p} [s]) > n(T)\cdot Z + R \right)  \right]  \\
&\leq \Pr \left[ \exists T \left( {\tt u}^T(\mathbb{B}^{\infty}_{k,q} [s]) > n(T)\cdot Z + R \right)  \right] \\ 
&= \Pr \left[{\tt st}^Z [\mathbb{B}_{k,q}^{\infty} (\textbf{s}) ] > R \right]  \\
&= \Pr \left[{\tt st} [\mathbb{B}_{k,q}^Z (\textbf{s}) ] > R \right]  
\end{align*}
Finally, to complete the argument, we use the following result from basic information theory:
\begin{lemma}
Let $X$ be a discrete random variable that takes on only non-negative integer values. Then
\begin{equation} \label{eqn:infobasic}
\mathbb{E}[X] = \sum\limits_{i=1}^{\infty} \Pr(X \geq i)
\end{equation}
\end{lemma}
Summing Eq.~\ref{eqn:infobasic} for $R = 0, 1, 2 \cdots$ (which corresponds to $R_1, R_2 \geq 1$) we get that,
\begin{align*}
\mathbb{E} [R_1] &= \sum_R \Pr \left[{\tt st} [\mathbb{A}_{k,p}^Z (\textbf{s}) ] > R \right] \\
&\leq \sum_R \Pr \left[{\tt st} [\mathbb{B}_{k,p}^Z (\textbf{s}) ] > R \right] \\
&= \mathbb{E} [R_2] 
\end{align*}
Which completes the proof for Theorem~\ref{thm:DPhelps}. $\blacksquare $\\

\textbf{Proof of Theorem~\ref{thm:stashbounds}: }Using Theorem~\ref{thm:DPhelps}, we know that the stash usage is ``lower'' for non-zero values of $\epsilon$. Hence, it suffices to give stash bounds when $\epsilon = 0$. As in the proof for Theorem~\ref{thm:DPhelps}, we conceptually extend the server storage to contain a complete binary tree with height $L$, where the sub-trees form the lower levels of the extended binary tree. We can see that for $\epsilon = 0$, the \ourprotocol{} protocol with the additional storage reduces to the Path ORAM protocol and hence the stash size of \ourprotocol{} can bounded as: 

\begin{equation}
\Pr \left[{\tt st} [\mathbb{O}_{k,p}^Z (\textbf{s}) ] > R + Z \cdot 2^k \right] \leq 14 (0.6002)^R
\end{equation}

This completes the proof of the stash bounds\footnote{Line 4 in Flush protocol in \ourprotocol{} ensures that buckets never store more than $Z$ blocks. Hence, capturing the stash failure probability suffices.}. $\blacksquare $\\

\textbf{Proof of Theorem~\ref{thm:bandwidth}: }The proof follows by noting that the depth of each sub-tree is equal to $(L + 1 - k)$ and hence number of blocks are transferred per access is $2Z$ times the depth. $\blacksquare $\\

\textbf{Proof of Theorem~\ref{thm:DPORAMtoPIR}: }The proof follows directly from the setup and the definition of DP-PIR. Given any two adversarial queries $Q_i, Q_j$ for database records, we consider these as ORAM input access sequences, each with only single access. Since these access sequences differ in a single access, for any output observation $O$:
\begin{equation}
\Pr [O|Q_i] \leq e^{\epsilon} \Pr [O|Q_j] + \delta
\end{equation}
which is the privacy guarantee for DP-PIR. $\blacksquare$\\

\textbf{Proof of Theorem~\ref{thm:DPPIRAnonymousChannels}: }$\delta$ captures the failure probability of our system and hence we can union bound this failure probability across the $u$ users. Across $u$ users, the failure probability $\delta'$ can be bounded as: $\delta' \leq u \cdot \delta$ ($\delta'  \leq 1$). With probability $1-\delta'$, the composite system is now $\epsilon$-differentially private and we can use the Composition Lemma\footnote{Since $u \gg 1$, the law of large numbers holds for the proof of the Composition Lemma from Toledo \etal~\cite{goldbergDPPIR}.} to get the following bound on $\epsilon'$:

\begin{align*}
\epsilon' &= \ln \left(e^{2\epsilon} + u - 1 \right) - \ln u \\
&= \ln \left( \frac{e^{2\epsilon}}{u} + \frac{u - 1}{u}\right) \\
&\leq \ln \left( \frac{e^{2\epsilon}}{u} + 1 \right) \leq \frac{e^{2\epsilon}}{u}  
\end{align*}

Where the last two inequalities follow from (1) $u \gg 1$ (2) $u \gg e^{2\epsilon}$ and $\ln (1 + x) \leq x$ when $x > -1$. Finally, since the Composition Lemma from Toledo \etal~\cite{goldbergDPPIR} holds with a failure probability $1 - neg(u)$, we can incorporate this failure into the delta bound (union bound) to complete the proof of Theorem~\ref{thm:DPPIRAnonymousChannels}. $\blacksquare$\\

\textbf{Proof of Theorem~\ref{thm:DPPIRComposition}: }The proof follows naturally from the results of Theorem~\ref{thm:composability}. DP-PIR protocol relies on a DP-ORAM protocol and hence the privacy of $m$ $(\epsilon, \delta)$-DP-PIR queries sent to the trusted hardware by Theorem~\ref{thm:composability} can be bound by a single DP-PIR protocol with privacy $(m\epsilon, m\delta)$-DP-PIR guarantees. $\blacksquare$

\begin{figure}[!tp]
\centering
\includegraphics[width=0.9\linewidth]{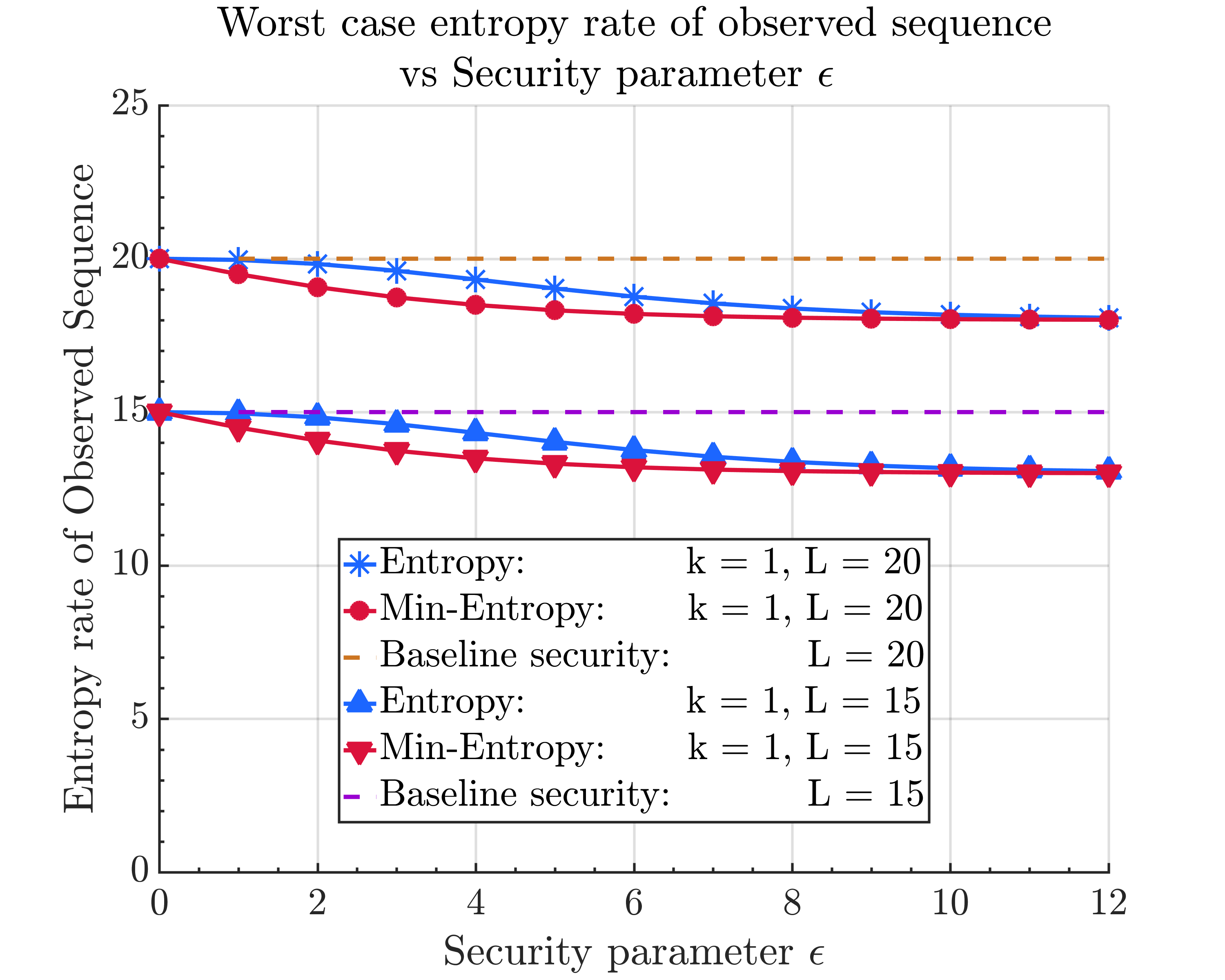}
\vspace{-5pt}
\caption{\textbf{Entropy rate for worst case access sequence for $L = 20$ and $15$ and $k = 1$.}}
\label{fig:entropyrate}
\end{figure}

\section{Entropy Calculation for DP}\label{appendix:DPdefense}
Next, we provide an interpretation of the privacy guarantees of the protocol in terms of entropy~\cite{paulDP}. Specifically, we find the worst case entropy of the observed access sequence for any given input access sequence. This entropy reflects the adversaries' uncertainty in the observed access sequence given any input sequence. We compute this entropy as follows:

Let $X_1, X_2, \hdots , X_M$ denote the random variables indicating the accessed location for the given access sequence $\textbf{a}$ of logical block addresses (i.e., $X_i$ is the random variable for $y_i$ where $\textbf{y} = \{ y_i \}_{i=1}^M$ as in Section~\ref{sec:analysis}). Let $\pi(\cdot)$ denote a function which maps an index in $\{1,2, \hdots , M\}$ to the location of the previous access of the same data block. 
In other words, if some data block $a$ was accessed at the $i^{{\tiny \mbox{th}}}$ $(x_i = a)$ and then at the $j^{{\tiny \mbox{th}}}$ $(x_j = a)$ location, then $\pi(j) = i$.\footnote{Formally, we can define $\pi(\cdot): \{1,2, \hdots , M\} \to \{1,2, \hdots , M-1\} \cup \phi $ as $\pi(j) = \max i$ s.t. $x_i = x_j$ and $i < j$ and $\phi$ otherwise.
\begin{equation*}
\pi(j) = 
\begin{cases}
\max i & \text{s.t. $x_i = x_j$ and $i < j$}\\
\phi	& \text{otherwise}
\end{cases}
\end{equation*}
}

Let $H(\cdot)$ denote the Shannon Entropy. If we have perfect security, $H(X_1, X_2, \hdots , X_M) = M \cdot \log N$ and 
hence the \textit{entropy rate} is $\log N$ (entropy per access). Using DP-ORAM reduces this entropy and we compute this loss in entropy below. We know that $P(X_i) = \textrm{Uniform $(\{0,1,2, \hdots , 2^L-1 \})$}$ if $\pi(i) = \phi$ and $P(X_i | X_{\pi(i)}) = D$, where $D$ is the distribution specified in Eq.~\ref{eqn:distributionD}. Hence, we can compute the entropy of the complete sequence as follows:
\begin{align*}
H(X_1, X_2, \hdots X_M) &= H(X_1) + \sum\limits_{i=2}^M H(X_i | X_{i-1} \hdots X_1) \\
&\geq H(X_1) + \sum\limits_{i=2}^M H(X_i | X_{i-1} \hdots X_1, \pi(i)) \\
&= H(X_1) + \sum\limits_{i=2}^M H(X_i | X_{\pi(i)}) \\
&= \alpha \cdot \log N + (M-\alpha) \cdot H(D) \\
&\geq M \cdot H(D)
\end{align*}
Where $\alpha$ is the number of accesses such that $\pi(\cdot) = \phi$ i.e., accessed for the first time. We know that $1 \leq \alpha \leq N$ and the entropy rate is $\frac{1}{M} H(X_1, X_2, \hdots H_M) \geq H(D)$.

For the chosen distribution $D$, we can compute $H(D)$ as:
\begin{equation}\label{eq:HofD}
H(D) = - (N-2^{L-k}) p_{{\tiny \mbox{min}}} \log p_{{\tiny \mbox{min}}} - 2^{L-k} p_{{\tiny \mbox{max}}} \log p_{{\tiny \mbox{max}}}
\end{equation}
since there are $N-2^{L-k}$ leaves with probability $p_{{\tiny \mbox{min}}}$ and $2^{L-k}$ leaves with probability $p_{{\tiny \mbox{max}}}$. The entropy rate is hence lower bounded by the expression in Eq.~\ref{eq:HofD}. In a similar manner, we compute the min-entropy of the observed sequence $H_{\infty} (X_1, X_2, \hdots , X_M)$\footnote{Min-entropy is defined as $H_{\infty}(X) = - \log \max\limits_{i} p(x_i)$.}.
\begin{align} \label{eqn:MinH}
H_{\infty}(X_1, X_2, \hdots H_M) &= - \log \left[ (1/N)^{\alpha} \cdot p_{{\tiny \mbox{max}}}^{M-\alpha} \right] \nonumber \\
&\geq - M \cdot \log p_{{\tiny \mbox{max}}}
\end{align}

Fig.~\ref{fig:entropyrate} plots the lower bound on the entropy rate of the observed access sequence as a function of the $\epsilon$ (numerically). We can see that the entropy rate decreases as $\epsilon$ increases but the decrease is small for moderate values of $\epsilon$. For instance, for $L=20$, using $\epsilon = 3$ results in a loss of roughly $0.3$-bits of entropy and for $\epsilon = 2$ the loss is $0.1$ bits (contrast this with the performance improvements presented in Section~\ref{subsec:evalresults} and Fig.~\ref{StashSecurity}). Even at larger values of $\epsilon$, the loss is less than $2$-bits.

The entropy loss analysis bounds the \textit{reduction in the entropy of the next access observed by the adversary given access to an arbitrary number of previous accesses (hence the asymptotic analysis).} This can be used to argue about the entropy loss of the entire sequence as a function of the number of accesses (using the bounds from Fig.~\ref{fig:entropyrate} and Eq.~\ref{eq:HofD},\ref{eqn:MinH}). In other words, given a number of accesses $M$, the baseline of perfect security would guarantee an uncertainty of the observed sequence to be $M \cdot \log N = 20 \cdot M$-bits (considering $L = \log N = 20$) whereas using DP-ORAM will reduce this entropy to $M \times$entropy per access. To put these in perspective, for an access sequence of length $10^3$ the entropy loss for $\epsilon = 2$ will reduce the entropy of the observed sequence from 20,000-bits to 19,900-bits (by about $0.5$\%).

%
%
%
%
\section{Advanced Composition theorem - DP}\label{appendix:advancedcomposition}
Advanced composition theorem bounds the privacy guarantees of a mechanism under \textit{$k$-fold adaptive composition} (for details refer to Chapter 3 of~\cite{differentialprivacy}). Formally it is stated as: 

\begin{theorem}[\textbf{Advanced composition}]\label{thm:advancedcomposition}
\textbf{For all $\epsilon, \delta, \delta' \geq 0$, the class of $(\epsilon, \delta)$-differentially private mechanisms satisfies $(\epsilon', k\delta + \delta')$-differential privacy under k-fold adaptive composition for:
\begin{equation}\label{eqn:advancedcomposition}
\epsilon' = \sqrt{2k \ln (1/\delta')} \epsilon + k\epsilon (e^{\epsilon} - 1)
\end{equation}}
\end{theorem}

The above result provides a guideline for setting the privacy budget under composition. Using the advanced composition theorem, we can see that the privacy budget scales sub-linearly with the number of queries (Example 3.7 in~\cite{differentialprivacy}).


\end{document}